\documentclass[aps,prl,twocolumn,amsmath,amssymb,nofootinbib,superscriptaddress,floatfix,reprint,longbibliography]{revtex4-1}
\usepackage[dvips]{graphicx}
\usepackage{latexsym}
\usepackage{amsmath}
\usepackage{amsfonts}
\usepackage{amssymb}
\usepackage{bm}
\usepackage{color}
\usepackage{txfonts}
\usepackage{float}
\usepackage{braket}
\usepackage{url}
\usepackage{CJKutf8}
\usepackage[colorlinks=true, urlcolor=blue, linkcolor=blue, citecolor=blue, pdftex]{hyperref}
\usepackage{ulem}
\begin{document}
	\newcommand{\fig}[2]{\includegraphics[width=#1]{#2}}
	\def\gsim{~\rlap{$>$}{\lower 1.0ex\hbox{$\sim$}}}
	\setlength{\unitlength}{1mm}
	\newcommand{\bea}{\begin{equation} \begin{aligned}}
    \newcommand{\eea}{\end{aligned} \end{equation} }
    \def\abs#1{\left|{#1}\right|}          			
    \def\bra#1{\left<{#1}\right|}					
    \def\ket#1{\left|{#1}\right>}					

\title {Majorana zero modes under electron correlation}
\author{Ziyue Qi}
\affiliation{Beijing National Laboratory for Condensed Matter Physics and Institute of Physics,
	Chinese Academy of Sciences, Beijing 100190, China}
\affiliation{University of Chinese Academy of Sciences, Beijing 100190, China}

\author{Hongming Weng}
\email{hmweng@iphy.ac.cn}
\affiliation{Beijing National Laboratory for Condensed Matter Physics and Institute of Physics,
	Chinese Academy of Sciences, Beijing 100190, China}
\affiliation{University of Chinese Academy of Sciences, Beijing 100190, China}

\author{Kun Jiang}
\email{jiangkun@iphy.ac.cn}
\affiliation{Beijing National Laboratory for Condensed Matter Physics and Institute of Physics,
	Chinese Academy of Sciences, Beijing 100190, China}
\affiliation{University of Chinese Academy of Sciences, Beijing 100190, China}

\date{\today}

\begin{abstract}
In this work, we perform a systematic investigation of the correlated topological superconductors (TSCs), especially their non-trivial Majorana zero modes (MZMs). Compared to the non-interacting MZMs, the emerged correlated MZMs become projected MZMs. To prove this, we study the topological superconducting nanowire under the Hubbard and Hatsugai-Kohmoto interactions. Both of them become correlated TSCs under a magnetic field. Their topological properties are numerically computed by the Wilson loop and entanglement spectrum.
We successfully extract the projected MZMs connecting the ground state and excited state through exact diagonalization. We also extend our results to the spinful Kitaev chain. Our results can provide a new perspective and understanding of the correlated MZMs.
\end{abstract}
\maketitle

\textit{Introduction}
Majorana zero modes (MZMs) in condensed matter physics are quasiparticle extensions of Majorana fermions (MFs) in particle physics, where particles are their own antiparticles \cite{Kitaev_2001,Alicea_2012,Beenakker,Franz_RevModPhys.87.137}. 
Besides their value in fundamental science, MZMs also host a strong potential application for fault-tolerant quantum computing because of their non-Abelian exchange statistics \cite{Kitaev_2001,Beenakker}. 
However, the realization of MZMs remains a great challenge. Owing to the development of topological insulators and topological superconductors (TSCs) during the past twenty years \cite{Qi_RevModPhys.83.1057,Hasan_RevModPhys.82.3045}, 
tremendous efforts have been spent on the search for MFs in superconductors \cite{fu_kane_PhysRevLett.100.096407,Sau_PhysRevLett.104.040502,Sau_PhysRevLett.105.077001,Oreg_PhysRevLett.105.177002,iron-chain,Potter_PhysRevLett.105.227003}.
For example, by proximity coupled to an $s$-wave superconductor, a semiconductor nanowire with strong Rashba spin-orbit coupling (SOC) under external Zeeman field hosts two MZMs at its ends \cite{Sau_PhysRevLett.105.077001,Oreg_PhysRevLett.105.177002}. 
This proposal has been widely explored \cite{Kouwenhoven_nano12,nano_wire2,nano_wire_review}, becoming one of the promising candidates for realizing topological MZMs.

\begin{figure}[t]
	\begin{center}
		\fig{3.4in}{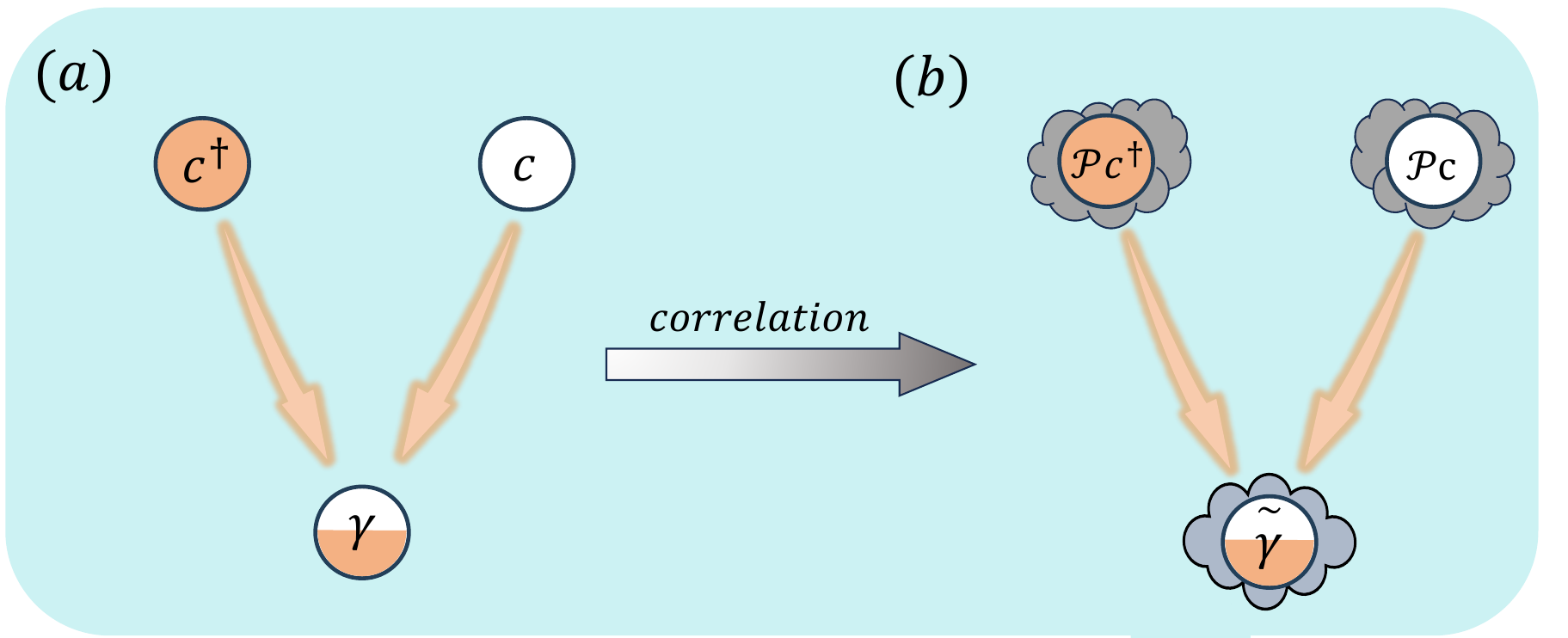}
		\caption{ A schematic diagram of Majorana operators without (a) and with (b) electron-electron correlation. Owing to correlation, the electron operators should be supplemented with the projection operator $\mathcal{P}$, which projects out all the high-energy Hilbert space.
        The non-interacting Majorana operator $\gamma$ will be replaced by its correlated version $\tilde{\gamma}$.
  \label{fig1}}
	\end{center}
\end{figure}

Formally, one Majorana operator $\gamma_\alpha$ with quantum index $\alpha$ can be written as the combination of fermion creation and annihilation operators $\gamma_\alpha=c^\dagger_\alpha+c_\alpha$, as illustrated in Fig.\ref{fig1}(a).
The above analysis and previous understanding are based on the non-interacting electron approach \cite{Sato_2017,Alicea_2012}. 
On the other hand, it is widely known that the electron-electron correlation plays a significant role in transition metal oxides, like the high-temperature superconductivity in doped cuprates. When this correlation effect is unavoidable, how this Majorana operator evolves becomes an interesting question.
In this work, we find a natural generalization MZM $\gamma_\alpha$ to correlated MZM $\tilde{\gamma}_\alpha=\hat{P}c^\dagger_\alpha+\hat{P}c_\alpha$, where $\hat{P}$ is the projection operator treating electron correlation, as illustrated in Fig.\ref{fig1}(b).




\textit{Nanowire}
To demonstrate the above results, we start with the semiconductor superconducting nanowire example under electron correlation.
The nanowire Hamiltonian in the basis $\Psi^\dagger_{{k}} = \{c^\dagger_{{k},\uparrow},c^\dagger_{{k},\downarrow},c_{-{k},\uparrow},c_{-{k},\downarrow}\}$ can be written as
\bea
\label{nano-HK-k-ham}
H_{nano}=\sum_{{k}} \frac{1}{2}\Psi^\dagger_{{k}} h(k)\Psi_{{k}} + H_{int} 
\eea
where the $h(k)$ is the non-interacting part describing a superconducting nanowire under Zeeman field $h$ and $H_{int}$ is the interaction part. To simplify our discussion, we use a lattice version of $h(k)=\varepsilon_{{k}}\sigma_z s_0 + h \sigma_z s_z + \lambda \sin(k)\sigma_z s_y - \Delta \sigma_y s_y$ \cite{Sau_PhysRevLett.105.077001,Oreg_PhysRevLett.105.177002}. Here, $s_i$/$\sigma_i$ ($i=0,x,y,z$) are the Pauli matrix for spin/particle-hole degree of freedom. $\varepsilon_{{k}}=-2t\cos(k)+(\mu+2t)$, which corresponds to a parabolic dispersion around $k=0$. We use $t=1$ as the energy unit. $\Delta$ is the proximity superconducting pairing field and 
$\lambda$ is the Rashba SOC coupling strength.
This superconducting nanowire becomes a topological superconductor with MZMs, when $h>h_c=\sqrt{\mu^2+\Delta^2}$ without interaction ~\cite{Oreg_PhysRevLett.105.177002,Sau_PhysRevLett.105.077001}.

Turing on the interaction $H_{int}$, the problem normally becomes analytically unsolvable. To achieve a solvable limit, we first consider the Hatsugai-Kohmoto (HK) interaction $Un_{{k},\uparrow}n_{{k},\downarrow}$ with local interaction in reciprocal space \cite{HK}. We find that the MZM that emerged from HK interaction shows a similar structure as the Hubbard interaction $U n_{i\uparrow}n_{i\downarrow}$ with local interaction in the real space.
Recently, the HK model has won wide attention owing to its special role in uncovering the electron correlation effect from a solvable perspective. Non-Fermi liquid and unconventional superconductor\cite{Phillips_hk_doped_sc,phillips_PhysRevB.105.184509,Huang2022,Li_2022,li2023correlated,kunyang_PhysRevB.103.024529}, Kondo physics\cite{yang_10.1093/pnasnexus/pgad169,ch2021dilute,zhong_PhysRevB.106.155119} have been intensively studied via HK model.
HK model has also been extended to topological physics including correlated Chern insulators and topological Mott insulators \cite{mai_PhysRevResearch.5.013162,Mai2023,zhao_PhysRevLett.131.106601}

\begin{figure}[t]
	\begin{center}
		\fig{3.5in}{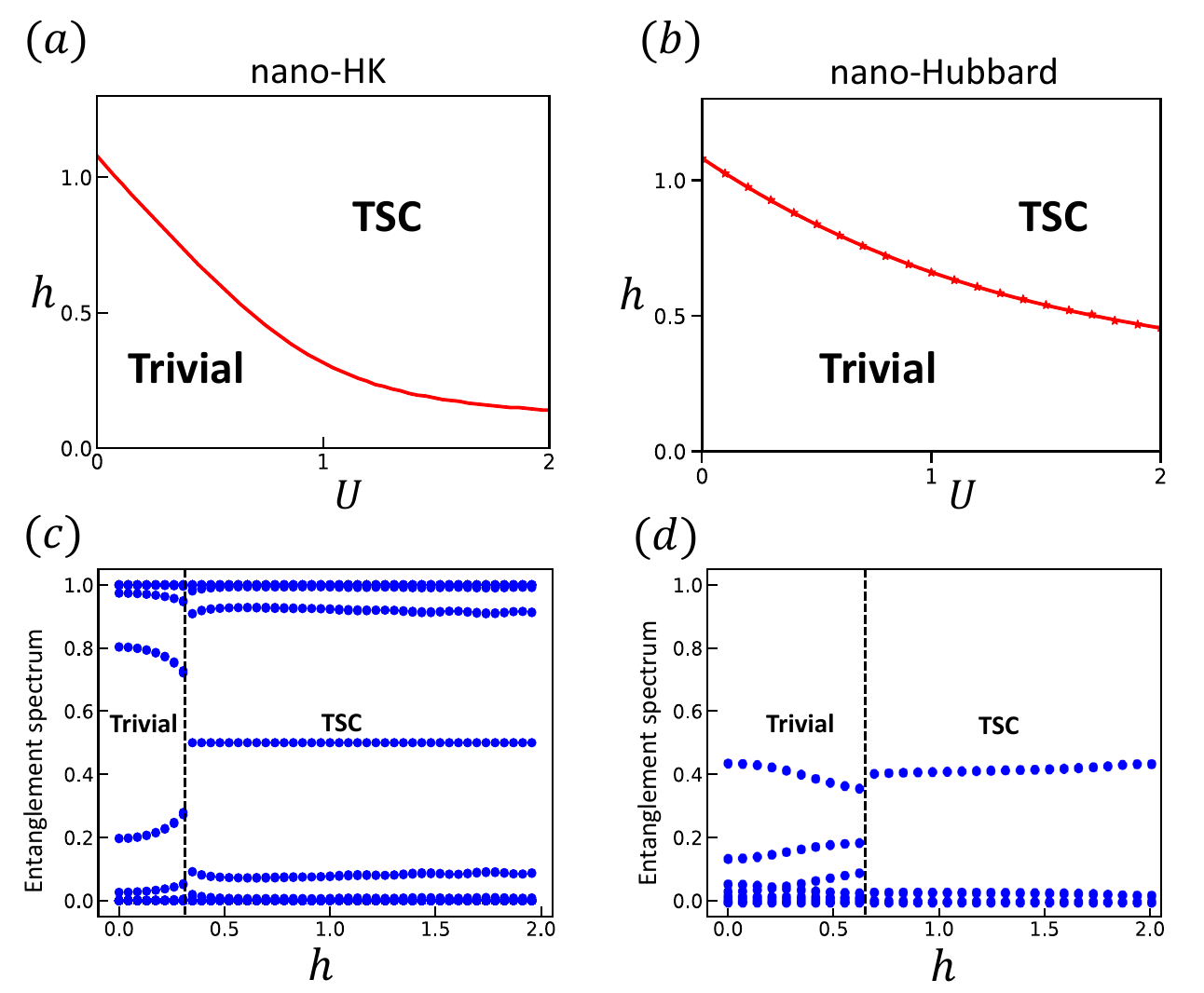}
		\caption{The phase diagrams of topological nanowire under HK (a) and Hubbard interaction (b) with $\lambda=0.5$, $\mu=-1$ and $\Delta=0.4$. The red lines are the phase transition lines between TSC and topological trivial phases. (a) is calculated in reciprocal space and (b) is calculated by ED with 12 sites under PBC. To prove the topological properties, we calculate the entanglement spectrum of the model under HK and Hubbard interaction along $U=1$ in (c) and (d). (c) is calculated explicitly based on $\ket{GS}_k$ and (d) is calculated by performing ED with 12 sites under PBC. The phase boundaries are separated by dash black lines.
			\label{fig2}}
	\end{center}
\end{figure}

\textit{phase-diagram}
Since HK interaction is local in $k$-space, we can easily diagonalize the $H$ using the Fock basis $\ket{n_{k,\uparrow},n_{-k,\uparrow},n_{k,\downarrow},n_{-k,\downarrow}}$ at each $k$ point, as listed in SM. Owing to this special property, any eigen-wavefunction can be written as a product state of eigen-wavefunction at each $k$ point. For example, the ground state wavefunction is $\ket{GS}=\prod_k\ket{GS}_k$.
Then, the many-body energy gap $\Delta_{gap}$ of $H$ is determined by the minimal gap at each $k$ point.
Through extracting $\Delta_{gap}$, we obtain the phase diagram of the HK model, as plotted in Fig.\ref{fig2}(a).
Similar to nanowires under the Zeeman field, the phase diagram is separated by two regions: trivial and TSC, where the dashed line with $\Delta_{gap}=0$ is the topological phase transition line. 
Here, we apply two methods to determine the topological index for each phase.
Owing to the fact that $\ket{GS}$ is a product state, the Zak phase calculation used in the non-interacting limit can be generalized to the many-body problem by calculating the Wilson loop \cite{wilson_loop,Zak_PhysRevLett.62.2747}, the entanglement spectrum calculation for free fermion system can be also generalized to the many-body problem by performing Fourier transformation on the $\ket{GS}_k$ and calculating the eigenvalues of the correlation matrix $G_A$ of subsystem A\cite{wu_PhysRevB.91.115118,fan_PhysRevB.94.165167,Sirker_2014,Peschel_2009,po_PhysRevResearch.2.033069}. The entanglement spectrum along $U=1$ of the HK model are plotted in Fig.\ref{fig2}(c). And both methods reach the same conclusions, more detailed derivations and results are listed in SM.
From Fig.\ref{fig2}(a), we also find the critical value for $h_c(U)$ keeps decreasing with increasing $U$.
This fact results from the effective chemical potential shifting due to $U$ increasing.


Besides the HK model, we also obtain the phase diagram of the Hubbard interaction in Fig.\ref{fig2}(b). We solve the Hubbard phase diagram by performing exact diagonalization (ED) on a finite size chain (chain size L up to 12) under periodic boundary condition (PBC) using Quspin package \cite{quspin_10.21468/SciPostPhys.2.1.003,quspin_10.21468/SciPostPhys.7.2.020}. The many-body gaps here are determined by the energy difference between the ground state and 1${st}$ excited state at L=12. A similar phase diagram between the trivial and TSC phases is plotted in Fig.\ref{fig2}(b). Although the transition line is from the finite-size ED calculation, the actual phase boundary shifts a little bit without qualitatively changing the following discussion.
To detect the topological index of the Hubbard model, we also solve the entanglement spectrum under PBC by performing ED, as plotted in Fig.\ref{fig2}(d). In this process, the finite size chain is divided into two parts from the middle, and entanglement spectrums are the eigenvalues of the reduced density matrix of one subsystem. The results exhibit similar topological properties to that in the HK model, and distinguish the trivial and TSC phases, more detailed discussion are listed in SM.

\textit{MZMs}
Now, bulk-edge correspondences in TSC guarantee the emergence of MZMs for the above model with open boundary conditions (OBCs). 
For the open boundary cases, we also need the help of many-body computation. Before detailed discussions, we can conjecture their forms from a large $U$ limit \cite{hubbard-3,fulde2012electron}. For the Hubbard model, it is well-known that the double occupancy quenches the low-energy dynamics of charge motions. Hence, the electron operators can be replaced by a projection operator $P_{j\sigma}^{Hub}=1-n_{j,\overline{\sigma}}$, whose structures are
$\tilde{c}^{Hub}_{j,\sigma}=(1-n_{j,\overline{\sigma}}) c_{j,\sigma}$ in the large $U$ limit.
Here, we focus on the electron filling less than half while the filling greater than half takes a similar form by replacing $(1-n_{j,\overline{\sigma}})$ to $n_{j,\overline{\sigma}}$.

For the HK model, the projection operator is now acting on the $k$ space with $P_{k}^{HK}=1-n_{k,\overline{\sigma}}$ with $\tilde{c}^{HK}_{k,\sigma}=(1-n_{k,\overline{\sigma}}) c_{k,\sigma}$. 
Through Fourier transformation on $\tilde{c}^{HK}_{j,\sigma}=\frac{1}{\sqrt{L}}\sum_{k,\sigma}\tilde{c}^{HK}_{k,\sigma}e^{-ikj}$, a new form is obtained $\tilde{c}^{HK}_{j,\sigma}=c_{j,\sigma}-\frac{1}{L}\sum_{j_1,j_2}c_{j_1,\sigma}\;c^\dagger_{j_1+j_2-j,\overline{\sigma}}\;c_{j_2,\overline{\sigma}}$.



\begin{figure}[t]
	\begin{center}
		\fig{3.4in}{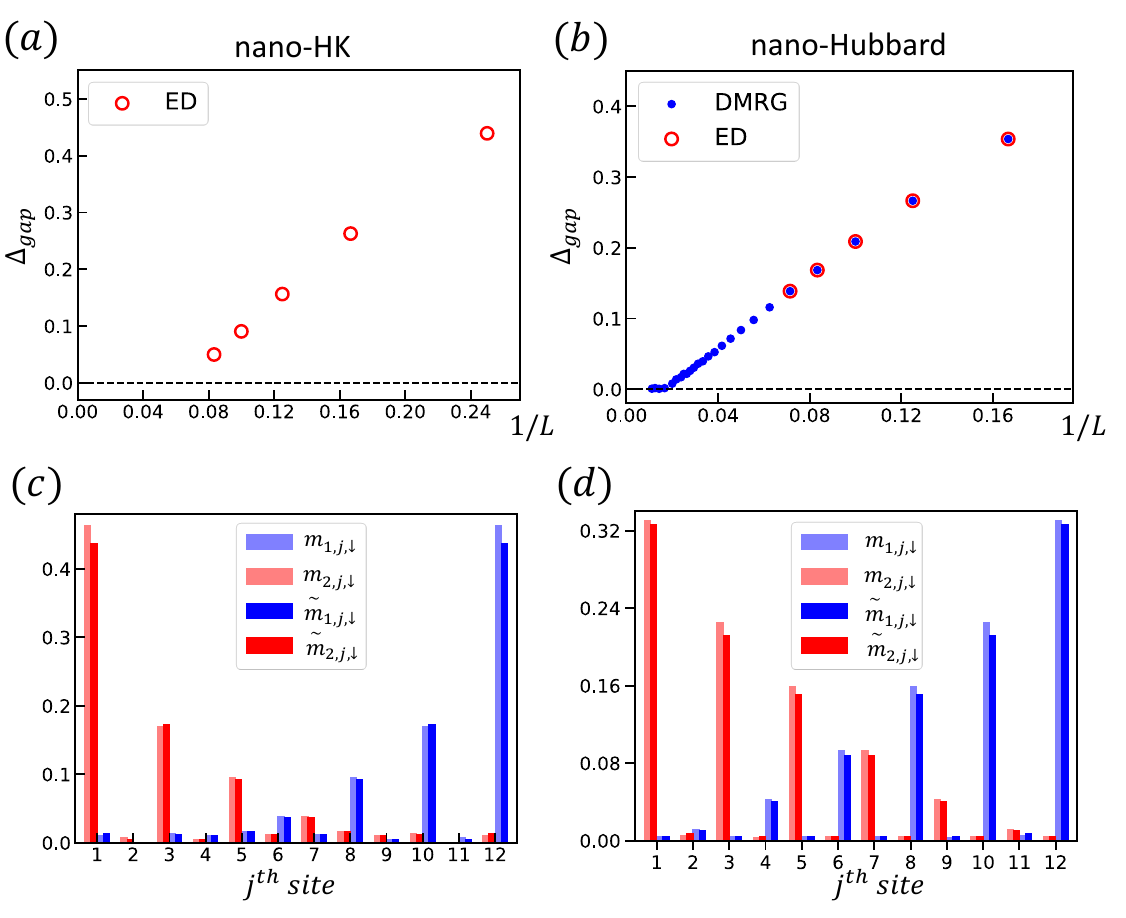}
		\caption{ The finite size extrapolations of the OBC nanowire models under HK (a) and Hubbard (b) interactions with $U=4, h=0.8, \mu=-1, \Delta=0.4, \lambda=0.5$. The ED calculations are obtained up to $L=12$. For the Hubbard model, we also apply a DMRG study up to $L=80$. Both of them show a gap-closing feature towards the long chain limit. Then, we extract the MZM excitations by comparing the distribution of $m_{\alpha,j,\downarrow}$ and $\tilde{m}_{\alpha,j,\downarrow}$ under HK (c) and Hubbard interactions (d). Both calculations are based on ED with $L=12$ with the same parameters in (a-b).
			\label{fig3}}
	\end{center}
\end{figure}

Then, we arrive at the central results of this work. The correlated Majorana operators can be written as the combination with correlated fermions $\tilde{c}^\dagger$ and $\tilde{c}$ with
\begin{eqnarray}
    \tilde{\gamma}_{1j,\sigma}&=&\tilde{c}_{j,\sigma}^\dagger+ \tilde{c}_{j,\sigma} \nonumber \\
    \tilde{\gamma}_{2j,\sigma}&=&i(\tilde{c}_{j,\sigma}^\dagger-\tilde{c}_{j,\sigma})
\end{eqnarray}
With the knowledge of correlated Majorana operators, we need to design a scheme extracting the MZM excitation from the numerical calculation. Owing to bulk-edge correspondences, the TSCs host gapless excitations in the thermodynamic limit. We check the energy gaps in the OBC conditions, as plotted in Fig.\ref{fig3} (a-b). The OBC gap in HK interaction goes towards zero more quickly than in the Hubbard cases up to $L=12$ ED. This feature relates to its infinite long-range interaction. We also apply the density matrix renormalization group (DMRG) calculations on the Hubbard chain using the ITensor package \cite{ITensor} and find it becoming gapless in the large L limit (Fig.\ref{fig3} (b)). Because of the non-trivial topological index, the gapless excitation must be the correlated MFs.

With the knowledge of projection operators and topological properties, we need to find the operator connecting the ground state and 1st excited state by $\hat{O}|GS\rangle=|ES\rangle$. This $\hat{O}$ operator must be composed of MZM $\gamma$ operators.
Hence, we calculate the probabilities of bare Majorana operator $\gamma$ without projection and the correlated Majorana $\tilde{\gamma}$ between $|GS\rangle$ and $|ES\rangle$, and make a comparison between them. We define the probabilities as follows
\bea
\label{probability}
m_{\alpha,j,\sigma}=|\bra{ES}\gamma_{\alpha,j,\sigma}\ket{GS}|^2\;,\;\tilde{m}_{\alpha,j,\sigma}=|\bra{ES}\tilde{\gamma}_{\alpha,j,\sigma}\ket{GS}|^2
\eea
with $\alpha=1,2$. These probabilities can be derived from Green function of MFs, which is discussed in SM. If MFs are purely from $\tilde{\gamma}$, it satisfies $m_{\alpha,j,\sigma} = \tilde{m}_{\alpha,j,\sigma}$.
Since $h$ is very large, we can focus on the spin-down components. The calculation results are plotted in Fig.\ref{fig3}(c-d). One can find that excitations are mostly localized at boundaries with the largest $m_1$ on the right and the largest $m_2$ on the left, as expected. Due to finite size calculation, the decay of MFs in the Hubbard model is relatively slower in the bulk. More importantly, we find that $m_{\alpha,j,\downarrow}$
almost equal to $\tilde{m}_{\alpha,j,\downarrow}$, which confirms the main component of correlated MFs are from correlated Majorana $\tilde{\gamma}$.

We want to add a note here. The projection operators we used here are only appropriate in the large $U$ limit, where double occupancy is strictly forbidden. For the numerical calculation, it will always contain small fluctuations away from a large $U$ limit and finite size. Hence, small deviations between $m_{\alpha,j,\sigma}$ and $ \tilde{m}_{\alpha,j,\sigma}$ are reasonable. For a finite $U$, one can replace the projection operator with the Gutzwiller projection $\Hat{P}_{G}$ \cite{Gutzwiller,Brinkman_PhysRevB.2.1324} or Kotliar-Ruckenstein slave boson approach \cite{KR-PhysRevLett.57.1362}, which calls for further exploration.



\begin{figure}[t]
	\begin{center}
		\fig{3.4in}{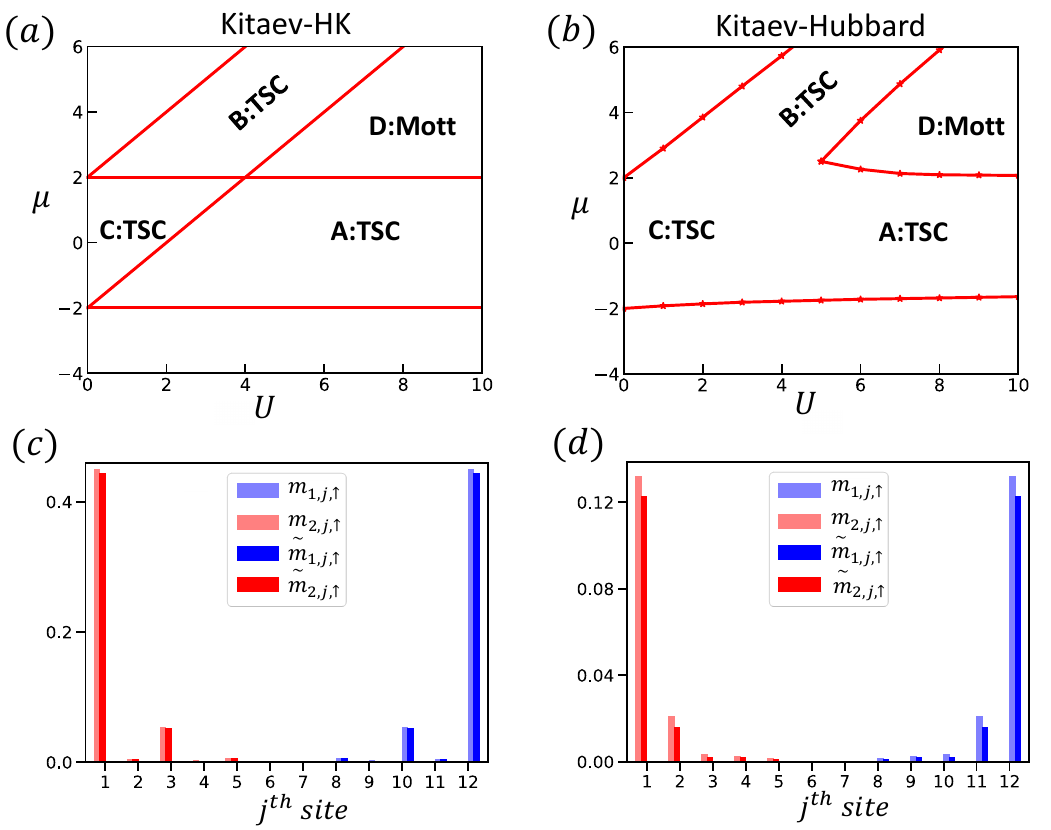}
		\caption{(a),(b) are the phase diagram of spinful Kitaev chain under HK and Hubbard interaction with $t=1,\Delta=0.4$. (a) is calculated in reciprocal space and (b) is calculated by ED under PBC with $L=12$, and red lines represent the phase transtion lines. We also extract the MZM excitations by comparing the distribution of $m_{\alpha,j,\uparrow}$ and $\tilde{m}_{\alpha,j,\uparrow}$ under HK (c) and Hubbard interactions (d) with $t=1,\Delta=0.4,U=16$ and $L=12$. 
			\label{fig4}}
	\end{center}
\end{figure}

\textit{spinful-kitaev-chain}
Furthermore, we want to generalize our results to other systems. Another simple system we consider here is the spinful Kitaev chain to encode electron-electron correlation. The Kitaev chain is the simplest model capturing the physics of TSC and MZMs \cite{Kitaev_2001}. The correlated spinful Kitaev chain is two copies of the spinless Kitaev chain with $p$-wave pairing under the basis $\Psi_{k,\sigma}^\dagger=\{c_{k,\sigma}^\dagger,c_{-k,\sigma}\}$,
\bea
\label{kitaev_hk-ham}
H_{Kitaev}=\sum_{k,\sigma}\frac{1}{2}\Psi_{k,\sigma}^\dagger h_\sigma(k) \Psi_{k,\sigma}+H_{int}
\eea
where $h_{\sigma}(k)=\varepsilon_k\sigma_z-\Delta_{\sigma,k}\sigma_y$ and $\varepsilon_k=-2t\cos(k)-\mu$, $\Delta_{\sigma,k}=(-1)^\sigma\cdot 2\Delta\sin(k)$ for the $p$-wave pairing term, and $H_{int}$ can be HK or Hubbard interaction,and the many-body Hamiltonian under HK interaction in $k$-space is listed in SM. The non-interacting spinful Kitaev chain contributes two MZMs at each chain end.

Similar to the analysis in the topological nanowire, we obtain the phase diagram of the above models by extracting the many-body gaps $\Delta_{gap}$, as plotted in Fig.\ref{fig4}. However, the phase diagrams of HK and Hubbard are obviously different here. Besides the conventional topological trivial phase, we find four phases for the HK model, as shown in Fig.\ref{fig4}(a). The phase C is the TSC phase with two MZMs at each chain end connecting with the non-interacting limit. 
Phase D is the Mott insulating phase with electron filling around half, which is topologically trivial. Besides these two, we find two additional phases A and B, which are particle-hole symmetric to each other along the line $\mu=\frac{U}{2}$. Both phases A and B are topologically nontrivial. 
To prove this, exact solutions of the topological index in the HK model, including the Zak phase and entanglement spectrums, are shown in SM. Since Zak phase is a $\mathbf{Z}_2$ index, its value equals $\pi$ in phases A and B, but $0$ in phase C. On the other hand, entanglement spectrums help us identify the accurate results with two non-trivial spectrums in A, B, and four in C, as plotted in SM.

The phase diagram of the Hubbard model becomes much simpler. Solving the Hubbard through ED, we 
obtain the phase diagram of the Hubbard model in Fig.\ref{fig4}(b). Besides the trivial phase, there are only two phases, phase C and the Mott phase.
To compare with the HK model, we also add two labels A and B in Fig.\ref{fig4}(b). However, these two regions are smoothly connected with phase-C without any gap-closing feature from ED. 
Numerical solutions of entanglement spectrums in the Hubbard model are also given in SM, which shows four non-trivial spectrums in regions A, B, and C. 
The form of Majorana operators is slightly different owing to the electron filling difference.  

Moving to the Majorana excitations, we check the bulk edge correspondences by calculating the open boundary many-body gaps numerically. The HK model and Hubbard models show similar gap-closing features as in the nanowire cases. Hence, we can believe the existence of topological excitations in the topological non-trivial regions.
Following the same strategy, we guess the projected MZMs and solve the probabilities defined in Eq.\ref{probability} in both HK and Hubbard models.
Notice that, the electron filling in region A/B is less/greater than half, so the projection operator of the electron in A and B is $(1-n_{\alpha,\sigma})$ and $n_{\alpha,\sigma}$ respectively, which leads to the different components of MFs and MZMs. Here, $\alpha$ is the index in real space for the Hubbard model and reciprocal space for the HK model. 
Taking the phase A as an example, the projection operator is mainly composed of $(1-n_{\alpha,\sigma})$. By comparing the probabilities between $\gamma$ and $\tilde{\gamma}$, we can find that the low-energy excitations are indeed from projection MZMs, where the probabilities are almost identical in the large $U$ limit, as shown in Fig.\ref{fig4}(c),(d). We also extract the projected MZMs in region B and double MZMs in region C, which are listed in SM.

\textit{summary}
In summary, we carry out a systematic investigation on the correlated TSCs and their non-trivial MZMs excitations. We find that the MZMs under correlation become projected MZMs under both HK and Hubbard interaction. To prove the projected MZMs under correlated TSCs, we studied two standard TSC models under HK and Hubbard interaction, which are topological superconducting nanowires under a magnetic field and spinful Kitaev chain. 

From their phase diagrams, there are always topological non-trivial regions under correlation. For the HK interaction, we have developed a general scheme to explicitly calculate its Zak phase and entanglement spectrum from their wavefunctions and the Wilson loop method. For the Hubbard interaction, the topological properties are obtained from the entanglement spectrum using ED. The bulk-edge correspondence guarantees the existence of gapless excitations for open boundary conditions.  We calculate the gap between the ground and 1st excited state of the OBC finite-size chains through ED and DMRG. The gap-closing features are achieved in the long chain limit. Furthermore, we successfully extract the projected MZMs connecting the ground state and excited state in both models by comparing the local excitation probabilities of unprojected and projected Majorana operators using ED in the large $U$ regime. Our findings provide a different perspective and understanding of MZMs under correlation. We hope our studies can stimulate the investigation of correlated TSCs.

\textit{Acknowledgement}
This work is supported by the Ministry of Science and Technology  (Grant No. 2022YFA1403901, No.2022YFA1403800), the National Natural Science Foundation of China (Grant No. NSFC-11888101, No. NSFC-12174428), the Strategic Priority Research Program of the Chinese Academy of Sciences (Grant No. XDB28000000, XDB33000000), the New Cornerstone Investigator Program, and the Chinese Academy of Sciences through the Youth Innovation Promotion Association (Grant No. 2022YSBR-048).

\bibliographystyle{apsrev4-1}
\bibliography{reference}

\begin{thebibliography}{45}%
\makeatletter
\providecommand \@ifxundefined [1]{%
 \@ifx{#1\undefined}
}%
\providecommand \@ifnum [1]{%
 \ifnum #1\expandafter \@firstoftwo
 \else \expandafter \@secondoftwo
 \fi
}%
\providecommand \@ifx [1]{%
 \ifx #1\expandafter \@firstoftwo
 \else \expandafter \@secondoftwo
 \fi
}%
\providecommand \natexlab [1]{#1}%
\providecommand \enquote  [1]{``#1''}%
\providecommand \bibnamefont  [1]{#1}%
\providecommand \bibfnamefont [1]{#1}%
\providecommand \citenamefont [1]{#1}%
\providecommand \href@noop [0]{\@secondoftwo}%
\providecommand \href [0]{\begingroup \@sanitize@url \@href}%
\providecommand \@href[1]{\@@startlink{#1}\@@href}%
\providecommand \@@href[1]{\endgroup#1\@@endlink}%
\providecommand \@sanitize@url [0]{\catcode `\\12\catcode `\$12\catcode `\&12\catcode `\#12\catcode `\^12\catcode `\_12\catcode `\%12\relax}%
\providecommand \@@startlink[1]{}%
\providecommand \@@endlink[0]{}%
\providecommand \url  [0]{\begingroup\@sanitize@url \@url }%
\providecommand \@url [1]{\endgroup\@href {#1}{\urlprefix }}%
\providecommand \urlprefix  [0]{URL }%
\providecommand \Eprint [0]{\href }%
\providecommand \doibase [0]{http://dx.doi.org/}%
\providecommand \selectlanguage [0]{\@gobble}%
\providecommand \bibinfo  [0]{\@secondoftwo}%
\providecommand \bibfield  [0]{\@secondoftwo}%
\providecommand \translation [1]{[#1]}%
\providecommand \BibitemOpen [0]{}%
\providecommand \bibitemStop [0]{}%
\providecommand \bibitemNoStop [0]{.\EOS\space}%
\providecommand \EOS [0]{\spacefactor3000\relax}%
\providecommand \BibitemShut  [1]{\csname bibitem#1\endcsname}%
\let\auto@bib@innerbib\@empty
\bibitem [{\citenamefont {Kitaev}(2001)}]{Kitaev_2001}%
  \BibitemOpen
  \bibfield  {author} {\bibinfo {author} {\bibfnamefont {A.~Y.}\ \bibnamefont {Kitaev}},\ }\href {\doibase 10.1070/1063-7869/44/10S/S29} {\bibfield  {journal} {\bibinfo  {journal} {Physics-Uspekhi}\ }\textbf {\bibinfo {volume} {44}},\ \bibinfo {pages} {131} (\bibinfo {year} {2001})}\BibitemShut {NoStop}%
\bibitem [{\citenamefont {Alicea}(2012)}]{Alicea_2012}%
  \BibitemOpen
  \bibfield  {author} {\bibinfo {author} {\bibfnamefont {J.}~\bibnamefont {Alicea}},\ }\href {\doibase 10.1088/0034-4885/75/7/076501} {\bibfield  {journal} {\bibinfo  {journal} {Reports on Progress in Physics}\ }\textbf {\bibinfo {volume} {75}},\ \bibinfo {pages} {076501} (\bibinfo {year} {2012})}\BibitemShut {NoStop}%
\bibitem [{\citenamefont {Beenakker}(2013)}]{Beenakker}%
  \BibitemOpen
  \bibfield  {author} {\bibinfo {author} {\bibfnamefont {C.}~\bibnamefont {Beenakker}},\ }\href {\doibase https://doi.org/10.1146/annurev-conmatphys-030212-184337} {\bibfield  {journal} {\bibinfo  {journal} {Annual Review of Condensed Matter Physics}\ }\textbf {\bibinfo {volume} {4}},\ \bibinfo {pages} {113} (\bibinfo {year} {2013})}\BibitemShut {NoStop}%
\bibitem [{\citenamefont {Elliott}\ and\ \citenamefont {Franz}(2015)}]{Franz_RevModPhys.87.137}%
  \BibitemOpen
  \bibfield  {author} {\bibinfo {author} {\bibfnamefont {S.~R.}\ \bibnamefont {Elliott}}\ and\ \bibinfo {author} {\bibfnamefont {M.}~\bibnamefont {Franz}},\ }\href {\doibase 10.1103/RevModPhys.87.137} {\bibfield  {journal} {\bibinfo  {journal} {Rev. Mod. Phys.}\ }\textbf {\bibinfo {volume} {87}},\ \bibinfo {pages} {137} (\bibinfo {year} {2015})}\BibitemShut {NoStop}%
\bibitem [{\citenamefont {Qi}\ and\ \citenamefont {Zhang}(2011)}]{Qi_RevModPhys.83.1057}%
  \BibitemOpen
  \bibfield  {author} {\bibinfo {author} {\bibfnamefont {X.-L.}\ \bibnamefont {Qi}}\ and\ \bibinfo {author} {\bibfnamefont {S.-C.}\ \bibnamefont {Zhang}},\ }\href {\doibase 10.1103/RevModPhys.83.1057} {\bibfield  {journal} {\bibinfo  {journal} {Rev. Mod. Phys.}\ }\textbf {\bibinfo {volume} {83}},\ \bibinfo {pages} {1057} (\bibinfo {year} {2011})}\BibitemShut {NoStop}%
\bibitem [{\citenamefont {Hasan}\ and\ \citenamefont {Kane}(2010)}]{Hasan_RevModPhys.82.3045}%
  \BibitemOpen
  \bibfield  {author} {\bibinfo {author} {\bibfnamefont {M.~Z.}\ \bibnamefont {Hasan}}\ and\ \bibinfo {author} {\bibfnamefont {C.~L.}\ \bibnamefont {Kane}},\ }\href {\doibase 10.1103/RevModPhys.82.3045} {\bibfield  {journal} {\bibinfo  {journal} {Rev. Mod. Phys.}\ }\textbf {\bibinfo {volume} {82}},\ \bibinfo {pages} {3045} (\bibinfo {year} {2010})}\BibitemShut {NoStop}%
\bibitem [{\citenamefont {Fu}\ and\ \citenamefont {Kane}(2008)}]{fu_kane_PhysRevLett.100.096407}%
  \BibitemOpen
  \bibfield  {author} {\bibinfo {author} {\bibfnamefont {L.}~\bibnamefont {Fu}}\ and\ \bibinfo {author} {\bibfnamefont {C.~L.}\ \bibnamefont {Kane}},\ }\href {\doibase 10.1103/PhysRevLett.100.096407} {\bibfield  {journal} {\bibinfo  {journal} {Phys. Rev. Lett.}\ }\textbf {\bibinfo {volume} {100}},\ \bibinfo {pages} {096407} (\bibinfo {year} {2008})}\BibitemShut {NoStop}%
\bibitem [{\citenamefont {Sau}\ \emph {et~al.}(2010)\citenamefont {Sau}, \citenamefont {Lutchyn}, \citenamefont {Tewari},\ and\ \citenamefont {Das~Sarma}}]{Sau_PhysRevLett.104.040502}%
  \BibitemOpen
  \bibfield  {author} {\bibinfo {author} {\bibfnamefont {J.~D.}\ \bibnamefont {Sau}}, \bibinfo {author} {\bibfnamefont {R.~M.}\ \bibnamefont {Lutchyn}}, \bibinfo {author} {\bibfnamefont {S.}~\bibnamefont {Tewari}}, \ and\ \bibinfo {author} {\bibfnamefont {S.}~\bibnamefont {Das~Sarma}},\ }\href {\doibase 10.1103/PhysRevLett.104.040502} {\bibfield  {journal} {\bibinfo  {journal} {Phys. Rev. Lett.}\ }\textbf {\bibinfo {volume} {104}},\ \bibinfo {pages} {040502} (\bibinfo {year} {2010})}\BibitemShut {NoStop}%
\bibitem [{\citenamefont {Lutchyn}\ \emph {et~al.}(2010)\citenamefont {Lutchyn}, \citenamefont {Sau},\ and\ \citenamefont {Das~Sarma}}]{Sau_PhysRevLett.105.077001}%
  \BibitemOpen
  \bibfield  {author} {\bibinfo {author} {\bibfnamefont {R.~M.}\ \bibnamefont {Lutchyn}}, \bibinfo {author} {\bibfnamefont {J.~D.}\ \bibnamefont {Sau}}, \ and\ \bibinfo {author} {\bibfnamefont {S.}~\bibnamefont {Das~Sarma}},\ }\href {\doibase 10.1103/PhysRevLett.105.077001} {\bibfield  {journal} {\bibinfo  {journal} {Phys. Rev. Lett.}\ }\textbf {\bibinfo {volume} {105}},\ \bibinfo {pages} {077001} (\bibinfo {year} {2010})}\BibitemShut {NoStop}%
\bibitem [{\citenamefont {Oreg}\ \emph {et~al.}(2010)\citenamefont {Oreg}, \citenamefont {Refael},\ and\ \citenamefont {von Oppen}}]{Oreg_PhysRevLett.105.177002}%
  \BibitemOpen
  \bibfield  {author} {\bibinfo {author} {\bibfnamefont {Y.}~\bibnamefont {Oreg}}, \bibinfo {author} {\bibfnamefont {G.}~\bibnamefont {Refael}}, \ and\ \bibinfo {author} {\bibfnamefont {F.}~\bibnamefont {von Oppen}},\ }\href {\doibase 10.1103/PhysRevLett.105.177002} {\bibfield  {journal} {\bibinfo  {journal} {Phys. Rev. Lett.}\ }\textbf {\bibinfo {volume} {105}},\ \bibinfo {pages} {177002} (\bibinfo {year} {2010})}\BibitemShut {NoStop}%
\bibitem [{\citenamefont {Nadj-Perge}\ \emph {et~al.}(2014)\citenamefont {Nadj-Perge}, \citenamefont {Drozdov}, \citenamefont {Li}, \citenamefont {Chen}, \citenamefont {Jeon}, \citenamefont {Seo}, \citenamefont {MacDonald}, \citenamefont {Bernevig},\ and\ \citenamefont {Yazdani}}]{iron-chain}%
  \BibitemOpen
  \bibfield  {author} {\bibinfo {author} {\bibfnamefont {S.}~\bibnamefont {Nadj-Perge}}, \bibinfo {author} {\bibfnamefont {I.~K.}\ \bibnamefont {Drozdov}}, \bibinfo {author} {\bibfnamefont {J.}~\bibnamefont {Li}}, \bibinfo {author} {\bibfnamefont {H.}~\bibnamefont {Chen}}, \bibinfo {author} {\bibfnamefont {S.}~\bibnamefont {Jeon}}, \bibinfo {author} {\bibfnamefont {J.}~\bibnamefont {Seo}}, \bibinfo {author} {\bibfnamefont {A.~H.}\ \bibnamefont {MacDonald}}, \bibinfo {author} {\bibfnamefont {B.~A.}\ \bibnamefont {Bernevig}}, \ and\ \bibinfo {author} {\bibfnamefont {A.}~\bibnamefont {Yazdani}},\ }\href {\doibase 10.1126/science.1259327} {\bibfield  {journal} {\bibinfo  {journal} {Science}\ }\textbf {\bibinfo {volume} {346}},\ \bibinfo {pages} {602} (\bibinfo {year} {2014})}\BibitemShut {NoStop}%
\bibitem [{\citenamefont {Potter}\ and\ \citenamefont {Lee}(2010)}]{Potter_PhysRevLett.105.227003}%
  \BibitemOpen
  \bibfield  {author} {\bibinfo {author} {\bibfnamefont {A.~C.}\ \bibnamefont {Potter}}\ and\ \bibinfo {author} {\bibfnamefont {P.~A.}\ \bibnamefont {Lee}},\ }\href {\doibase 10.1103/PhysRevLett.105.227003} {\bibfield  {journal} {\bibinfo  {journal} {Phys. Rev. Lett.}\ }\textbf {\bibinfo {volume} {105}},\ \bibinfo {pages} {227003} (\bibinfo {year} {2010})}\BibitemShut {NoStop}%
\bibitem [{\citenamefont {Mourik}\ \emph {et~al.}(2012)\citenamefont {Mourik}, \citenamefont {Zuo}, \citenamefont {Frolov}, \citenamefont {Plissard}, \citenamefont {Bakkers},\ and\ \citenamefont {Kouwenhoven}}]{Kouwenhoven_nano12}%
  \BibitemOpen
  \bibfield  {author} {\bibinfo {author} {\bibfnamefont {V.}~\bibnamefont {Mourik}}, \bibinfo {author} {\bibfnamefont {K.}~\bibnamefont {Zuo}}, \bibinfo {author} {\bibfnamefont {S.~M.}\ \bibnamefont {Frolov}}, \bibinfo {author} {\bibfnamefont {S.~R.}\ \bibnamefont {Plissard}}, \bibinfo {author} {\bibfnamefont {E.~P. A.~M.}\ \bibnamefont {Bakkers}}, \ and\ \bibinfo {author} {\bibfnamefont {L.~P.}\ \bibnamefont {Kouwenhoven}},\ }\href {\doibase 10.1126/science.1222360} {\bibfield  {journal} {\bibinfo  {journal} {Science}\ }\textbf {\bibinfo {volume} {336}},\ \bibinfo {pages} {1003} (\bibinfo {year} {2012})}\BibitemShut {NoStop}%
\bibitem [{\citenamefont {Deng}\ \emph {et~al.}(2016)\citenamefont {Deng}, \citenamefont {Vaitiekėnas}, \citenamefont {Hansen}, \citenamefont {Danon}, \citenamefont {Leijnse}, \citenamefont {Flensberg}, \citenamefont {Nygård}, \citenamefont {Krogstrup},\ and\ \citenamefont {Marcus}}]{nano_wire2}%
  \BibitemOpen
  \bibfield  {author} {\bibinfo {author} {\bibfnamefont {M.~T.}\ \bibnamefont {Deng}}, \bibinfo {author} {\bibfnamefont {S.}~\bibnamefont {Vaitiekėnas}}, \bibinfo {author} {\bibfnamefont {E.~B.}\ \bibnamefont {Hansen}}, \bibinfo {author} {\bibfnamefont {J.}~\bibnamefont {Danon}}, \bibinfo {author} {\bibfnamefont {M.}~\bibnamefont {Leijnse}}, \bibinfo {author} {\bibfnamefont {K.}~\bibnamefont {Flensberg}}, \bibinfo {author} {\bibfnamefont {J.}~\bibnamefont {Nygård}}, \bibinfo {author} {\bibfnamefont {P.}~\bibnamefont {Krogstrup}}, \ and\ \bibinfo {author} {\bibfnamefont {C.~M.}\ \bibnamefont {Marcus}},\ }\href {\doibase 10.1126/science.aaf3961} {\bibfield  {journal} {\bibinfo  {journal} {Science}\ }\textbf {\bibinfo {volume} {354}},\ \bibinfo {pages} {1557} (\bibinfo {year} {2016})}\BibitemShut {NoStop}%
\bibitem [{\citenamefont {Lutchyn}\ \emph {et~al.}(2018)\citenamefont {Lutchyn}, \citenamefont {Bakkers}, \citenamefont {Kouwenhoven}, \citenamefont {Krogstrup}, \citenamefont {Marcus},\ and\ \citenamefont {Oreg}}]{nano_wire_review}%
  \BibitemOpen
  \bibfield  {author} {\bibinfo {author} {\bibfnamefont {R.~M.}\ \bibnamefont {Lutchyn}}, \bibinfo {author} {\bibfnamefont {E.~P. A.~M.}\ \bibnamefont {Bakkers}}, \bibinfo {author} {\bibfnamefont {L.~P.}\ \bibnamefont {Kouwenhoven}}, \bibinfo {author} {\bibfnamefont {P.}~\bibnamefont {Krogstrup}}, \bibinfo {author} {\bibfnamefont {C.~M.}\ \bibnamefont {Marcus}}, \ and\ \bibinfo {author} {\bibfnamefont {Y.}~\bibnamefont {Oreg}},\ }\href {\doibase 10.1038/s41578-018-0003-1} {\bibfield  {journal} {\bibinfo  {journal} {Nature Reviews Materials}\ }\textbf {\bibinfo {volume} {3}},\ \bibinfo {pages} {52} (\bibinfo {year} {2018})}\BibitemShut {NoStop}%
\bibitem [{\citenamefont {Sato}\ and\ \citenamefont {Ando}(2017)}]{Sato_2017}%
  \BibitemOpen
  \bibfield  {author} {\bibinfo {author} {\bibfnamefont {M.}~\bibnamefont {Sato}}\ and\ \bibinfo {author} {\bibfnamefont {Y.}~\bibnamefont {Ando}},\ }\href {\doibase 10.1088/1361-6633/aa6ac7} {\bibfield  {journal} {\bibinfo  {journal} {Reports on Progress in Physics}\ }\textbf {\bibinfo {volume} {80}},\ \bibinfo {pages} {076501} (\bibinfo {year} {2017})}\BibitemShut {NoStop}%
\bibitem [{\citenamefont {Hatsugai}\ and\ \citenamefont {Kohmoto}(1992)}]{HK}%
  \BibitemOpen
  \bibfield  {author} {\bibinfo {author} {\bibfnamefont {Y.}~\bibnamefont {Hatsugai}}\ and\ \bibinfo {author} {\bibfnamefont {M.}~\bibnamefont {Kohmoto}},\ }\href {\doibase 10.1143/JPSJ.61.2056} {\bibfield  {journal} {\bibinfo  {journal} {Journal of the Physical Society of Japan}\ }\textbf {\bibinfo {volume} {61}},\ \bibinfo {pages} {2056} (\bibinfo {year} {1992})}\BibitemShut {NoStop}%
\bibitem [{\citenamefont {Phillips}\ \emph {et~al.}(2020)\citenamefont {Phillips}, \citenamefont {Yeo},\ and\ \citenamefont {Huang}}]{Phillips_hk_doped_sc}%
  \BibitemOpen
  \bibfield  {author} {\bibinfo {author} {\bibfnamefont {P.~W.}\ \bibnamefont {Phillips}}, \bibinfo {author} {\bibfnamefont {L.}~\bibnamefont {Yeo}}, \ and\ \bibinfo {author} {\bibfnamefont {E.~W.}\ \bibnamefont {Huang}},\ }\href {\doibase 10.1038/s41567-020-0988-4} {\bibfield  {journal} {\bibinfo  {journal} {Nature Physics}\ }\textbf {\bibinfo {volume} {16}},\ \bibinfo {pages} {1175} (\bibinfo {year} {2020})}\BibitemShut {NoStop}%
\bibitem [{\citenamefont {Zhao}\ \emph {et~al.}(2022)\citenamefont {Zhao}, \citenamefont {Yeo}, \citenamefont {Huang},\ and\ \citenamefont {Phillips}}]{phillips_PhysRevB.105.184509}%
  \BibitemOpen
  \bibfield  {author} {\bibinfo {author} {\bibfnamefont {J.}~\bibnamefont {Zhao}}, \bibinfo {author} {\bibfnamefont {L.}~\bibnamefont {Yeo}}, \bibinfo {author} {\bibfnamefont {E.~W.}\ \bibnamefont {Huang}}, \ and\ \bibinfo {author} {\bibfnamefont {P.~W.}\ \bibnamefont {Phillips}},\ }\href {\doibase 10.1103/PhysRevB.105.184509} {\bibfield  {journal} {\bibinfo  {journal} {Phys. Rev. B}\ }\textbf {\bibinfo {volume} {105}},\ \bibinfo {pages} {184509} (\bibinfo {year} {2022})}\BibitemShut {NoStop}%
\bibitem [{\citenamefont {Huang}\ \emph {et~al.}(2022)\citenamefont {Huang}, \citenamefont {Nave},\ and\ \citenamefont {Phillips}}]{Huang2022}%
  \BibitemOpen
  \bibfield  {author} {\bibinfo {author} {\bibfnamefont {E.~W.}\ \bibnamefont {Huang}}, \bibinfo {author} {\bibfnamefont {G.~L.}\ \bibnamefont {Nave}}, \ and\ \bibinfo {author} {\bibfnamefont {P.~W.}\ \bibnamefont {Phillips}},\ }\href {\doibase 10.1038/s41567-022-01529-8} {\bibfield  {journal} {\bibinfo  {journal} {Nature Physics}\ }\textbf {\bibinfo {volume} {18}},\ \bibinfo {pages} {511} (\bibinfo {year} {2022})}\BibitemShut {NoStop}%
\bibitem [{\citenamefont {Li}\ \emph {et~al.}(2022)\citenamefont {Li}, \citenamefont {Mishra}, \citenamefont {Zhou},\ and\ \citenamefont {Zhang}}]{Li_2022}%
  \BibitemOpen
  \bibfield  {author} {\bibinfo {author} {\bibfnamefont {Y.}~\bibnamefont {Li}}, \bibinfo {author} {\bibfnamefont {V.}~\bibnamefont {Mishra}}, \bibinfo {author} {\bibfnamefont {Y.}~\bibnamefont {Zhou}}, \ and\ \bibinfo {author} {\bibfnamefont {F.-C.}\ \bibnamefont {Zhang}},\ }\href {\doibase 10.1088/1367-2630/ac9548} {\bibfield  {journal} {\bibinfo  {journal} {New Journal of Physics}\ }\textbf {\bibinfo {volume} {24}},\ \bibinfo {pages} {103019} (\bibinfo {year} {2022})}\BibitemShut {NoStop}%
\bibitem [{\citenamefont {Li}\ \emph {et~al.}(2023)\citenamefont {Li}, \citenamefont {Jiang},\ and\ \citenamefont {Hu}}]{li2023correlated}%
  \BibitemOpen
  \bibfield  {author} {\bibinfo {author} {\bibfnamefont {P.}~\bibnamefont {Li}}, \bibinfo {author} {\bibfnamefont {K.}~\bibnamefont {Jiang}}, \ and\ \bibinfo {author} {\bibfnamefont {J.}~\bibnamefont {Hu}},\ }\href@noop {} {\enquote {\bibinfo {title} {Correlated bcs wavefunction approach to unconventional superconductors},}\ } (\bibinfo {year} {2023}),\ \Eprint {http://arxiv.org/abs/2309.02695} {arXiv:2309.02695 [cond-mat.supr-con]} \BibitemShut {NoStop}%
\bibitem [{\citenamefont {Yang}(2021)}]{kunyang_PhysRevB.103.024529}%
  \BibitemOpen
  \bibfield  {author} {\bibinfo {author} {\bibfnamefont {K.}~\bibnamefont {Yang}},\ }\href {\doibase 10.1103/PhysRevB.103.024529} {\bibfield  {journal} {\bibinfo  {journal} {Phys. Rev. B}\ }\textbf {\bibinfo {volume} {103}},\ \bibinfo {pages} {024529} (\bibinfo {year} {2021})}\BibitemShut {NoStop}%
\bibitem [{\citenamefont {Wang}\ and\ \citenamefont {Yang}(2023)}]{yang_10.1093/pnasnexus/pgad169}%
  \BibitemOpen
  \bibfield  {author} {\bibinfo {author} {\bibfnamefont {J.}~\bibnamefont {Wang}}\ and\ \bibinfo {author} {\bibfnamefont {Y.-f.}\ \bibnamefont {Yang}},\ }\href {\doibase 10.1093/pnasnexus/pgad169} {\bibfield  {journal} {\bibinfo  {journal} {PNAS Nexus}\ }\textbf {\bibinfo {volume} {2}},\ \bibinfo {pages} {pgad169} (\bibinfo {year} {2023})}\BibitemShut {NoStop}%
\bibitem [{\citenamefont {Setty}(2021)}]{ch2021dilute}%
  \BibitemOpen
  \bibfield  {author} {\bibinfo {author} {\bibfnamefont {C.}~\bibnamefont {Setty}},\ }\href@noop {} {\enquote {\bibinfo {title} {Dilute magnetic moments in an exactly solvable interacting host},}\ } (\bibinfo {year} {2021}),\ \Eprint {http://arxiv.org/abs/2105.15205} {arXiv:2105.15205 [cond-mat.str-el]} \BibitemShut {NoStop}%
\bibitem [{\citenamefont {Zhong}(2022)}]{zhong_PhysRevB.106.155119}%
  \BibitemOpen
  \bibfield  {author} {\bibinfo {author} {\bibfnamefont {Y.}~\bibnamefont {Zhong}},\ }\href {\doibase 10.1103/PhysRevB.106.155119} {\bibfield  {journal} {\bibinfo  {journal} {Phys. Rev. B}\ }\textbf {\bibinfo {volume} {106}},\ \bibinfo {pages} {155119} (\bibinfo {year} {2022})}\BibitemShut {NoStop}%
\bibitem [{\citenamefont {Mai}\ \emph {et~al.}(2023{\natexlab{a}})\citenamefont {Mai}, \citenamefont {Feldman},\ and\ \citenamefont {Phillips}}]{mai_PhysRevResearch.5.013162}%
  \BibitemOpen
  \bibfield  {author} {\bibinfo {author} {\bibfnamefont {P.}~\bibnamefont {Mai}}, \bibinfo {author} {\bibfnamefont {B.~E.}\ \bibnamefont {Feldman}}, \ and\ \bibinfo {author} {\bibfnamefont {P.~W.}\ \bibnamefont {Phillips}},\ }\href {\doibase 10.1103/PhysRevResearch.5.013162} {\bibfield  {journal} {\bibinfo  {journal} {Phys. Rev. Res.}\ }\textbf {\bibinfo {volume} {5}},\ \bibinfo {pages} {013162} (\bibinfo {year} {2023}{\natexlab{a}})}\BibitemShut {NoStop}%
\bibitem [{\citenamefont {Mai}\ \emph {et~al.}(2023{\natexlab{b}})\citenamefont {Mai}, \citenamefont {Zhao}, \citenamefont {Feldman},\ and\ \citenamefont {Phillips}}]{Mai2023}%
  \BibitemOpen
  \bibfield  {author} {\bibinfo {author} {\bibfnamefont {P.}~\bibnamefont {Mai}}, \bibinfo {author} {\bibfnamefont {J.}~\bibnamefont {Zhao}}, \bibinfo {author} {\bibfnamefont {B.~E.}\ \bibnamefont {Feldman}}, \ and\ \bibinfo {author} {\bibfnamefont {P.~W.}\ \bibnamefont {Phillips}},\ }\href {\doibase 10.1038/s41467-023-41465-6} {\bibfield  {journal} {\bibinfo  {journal} {Nature Communications}\ }\textbf {\bibinfo {volume} {14}},\ \bibinfo {pages} {5999} (\bibinfo {year} {2023}{\natexlab{b}})}\BibitemShut {NoStop}%
\bibitem [{\citenamefont {Zhao}\ \emph {et~al.}(2023)\citenamefont {Zhao}, \citenamefont {Mai}, \citenamefont {Bradlyn},\ and\ \citenamefont {Phillips}}]{zhao_PhysRevLett.131.106601}%
  \BibitemOpen
  \bibfield  {author} {\bibinfo {author} {\bibfnamefont {J.}~\bibnamefont {Zhao}}, \bibinfo {author} {\bibfnamefont {P.}~\bibnamefont {Mai}}, \bibinfo {author} {\bibfnamefont {B.}~\bibnamefont {Bradlyn}}, \ and\ \bibinfo {author} {\bibfnamefont {P.}~\bibnamefont {Phillips}},\ }\href {\doibase 10.1103/PhysRevLett.131.106601} {\bibfield  {journal} {\bibinfo  {journal} {Phys. Rev. Lett.}\ }\textbf {\bibinfo {volume} {131}},\ \bibinfo {pages} {106601} (\bibinfo {year} {2023})}\BibitemShut {NoStop}%
\bibitem [{\citenamefont {Yu}\ \emph {et~al.}(2011)\citenamefont {Yu}, \citenamefont {Qi}, \citenamefont {Bernevig}, \citenamefont {Fang},\ and\ \citenamefont {Dai}}]{wilson_loop}%
  \BibitemOpen
  \bibfield  {author} {\bibinfo {author} {\bibfnamefont {R.}~\bibnamefont {Yu}}, \bibinfo {author} {\bibfnamefont {X.~L.}\ \bibnamefont {Qi}}, \bibinfo {author} {\bibfnamefont {A.}~\bibnamefont {Bernevig}}, \bibinfo {author} {\bibfnamefont {Z.}~\bibnamefont {Fang}}, \ and\ \bibinfo {author} {\bibfnamefont {X.}~\bibnamefont {Dai}},\ }\href {\doibase 10.1103/PhysRevB.84.075119} {\bibfield  {journal} {\bibinfo  {journal} {Phys. Rev. B}\ }\textbf {\bibinfo {volume} {84}},\ \bibinfo {pages} {075119} (\bibinfo {year} {2011})}\BibitemShut {NoStop}%
\bibitem [{\citenamefont {Zak}(1989)}]{Zak_PhysRevLett.62.2747}%
  \BibitemOpen
  \bibfield  {author} {\bibinfo {author} {\bibfnamefont {J.}~\bibnamefont {Zak}},\ }\href {\doibase 10.1103/PhysRevLett.62.2747} {\bibfield  {journal} {\bibinfo  {journal} {Phys. Rev. Lett.}\ }\textbf {\bibinfo {volume} {62}},\ \bibinfo {pages} {2747} (\bibinfo {year} {1989})}\BibitemShut {NoStop}%
\bibitem [{\citenamefont {Wang}\ \emph {et~al.}(2015)\citenamefont {Wang}, \citenamefont {Xu}, \citenamefont {Wang},\ and\ \citenamefont {Wu}}]{wu_PhysRevB.91.115118}%
  \BibitemOpen
  \bibfield  {author} {\bibinfo {author} {\bibfnamefont {D.}~\bibnamefont {Wang}}, \bibinfo {author} {\bibfnamefont {S.}~\bibnamefont {Xu}}, \bibinfo {author} {\bibfnamefont {Y.}~\bibnamefont {Wang}}, \ and\ \bibinfo {author} {\bibfnamefont {C.}~\bibnamefont {Wu}},\ }\href {\doibase 10.1103/PhysRevB.91.115118} {\bibfield  {journal} {\bibinfo  {journal} {Phys. Rev. B}\ }\textbf {\bibinfo {volume} {91}},\ \bibinfo {pages} {115118} (\bibinfo {year} {2015})}\BibitemShut {NoStop}%
\bibitem [{\citenamefont {Ye}\ \emph {et~al.}(2016)\citenamefont {Ye}, \citenamefont {Mu},\ and\ \citenamefont {Fan}}]{fan_PhysRevB.94.165167}%
  \BibitemOpen
  \bibfield  {author} {\bibinfo {author} {\bibfnamefont {B.-T.}\ \bibnamefont {Ye}}, \bibinfo {author} {\bibfnamefont {L.-Z.}\ \bibnamefont {Mu}}, \ and\ \bibinfo {author} {\bibfnamefont {H.}~\bibnamefont {Fan}},\ }\href {\doibase 10.1103/PhysRevB.94.165167} {\bibfield  {journal} {\bibinfo  {journal} {Phys. Rev. B}\ }\textbf {\bibinfo {volume} {94}},\ \bibinfo {pages} {165167} (\bibinfo {year} {2016})}\BibitemShut {NoStop}%
\bibitem [{\citenamefont {Sirker}\ \emph {et~al.}(2014)\citenamefont {Sirker}, \citenamefont {Maiti}, \citenamefont {Konstantinidis},\ and\ \citenamefont {Sedlmayr}}]{Sirker_2014}%
  \BibitemOpen
  \bibfield  {author} {\bibinfo {author} {\bibfnamefont {J.}~\bibnamefont {Sirker}}, \bibinfo {author} {\bibfnamefont {M.}~\bibnamefont {Maiti}}, \bibinfo {author} {\bibfnamefont {N.~P.}\ \bibnamefont {Konstantinidis}}, \ and\ \bibinfo {author} {\bibfnamefont {N.}~\bibnamefont {Sedlmayr}},\ }\href {\doibase 10.1088/1742-5468/2014/10/P10032} {\bibfield  {journal} {\bibinfo  {journal} {Journal of Statistical Mechanics: Theory and Experiment}\ }\textbf {\bibinfo {volume} {2014}},\ \bibinfo {pages} {P10032} (\bibinfo {year} {2014})}\BibitemShut {NoStop}%
\bibitem [{\citenamefont {Peschel}\ and\ \citenamefont {Eisler}(2009)}]{Peschel_2009}%
  \BibitemOpen
  \bibfield  {author} {\bibinfo {author} {\bibfnamefont {I.}~\bibnamefont {Peschel}}\ and\ \bibinfo {author} {\bibfnamefont {V.}~\bibnamefont {Eisler}},\ }\href {\doibase 10.1088/1751-8113/42/50/504003} {\bibfield  {journal} {\bibinfo  {journal} {Journal of Physics A: Mathematical and Theoretical}\ }\textbf {\bibinfo {volume} {42}},\ \bibinfo {pages} {504003} (\bibinfo {year} {2009})}\BibitemShut {NoStop}%
\bibitem [{\citenamefont {Chang}\ \emph {et~al.}(2020)\citenamefont {Chang}, \citenamefont {You}, \citenamefont {Wen},\ and\ \citenamefont {Ryu}}]{po_PhysRevResearch.2.033069}%
  \BibitemOpen
  \bibfield  {author} {\bibinfo {author} {\bibfnamefont {P.-Y.}\ \bibnamefont {Chang}}, \bibinfo {author} {\bibfnamefont {J.-S.}\ \bibnamefont {You}}, \bibinfo {author} {\bibfnamefont {X.}~\bibnamefont {Wen}}, \ and\ \bibinfo {author} {\bibfnamefont {S.}~\bibnamefont {Ryu}},\ }\href {\doibase 10.1103/PhysRevResearch.2.033069} {\bibfield  {journal} {\bibinfo  {journal} {Phys. Rev. Res.}\ }\textbf {\bibinfo {volume} {2}},\ \bibinfo {pages} {033069} (\bibinfo {year} {2020})}\BibitemShut {NoStop}%
\bibitem [{\citenamefont {Weinberg}\ and\ \citenamefont {Bukov}(2017)}]{quspin_10.21468/SciPostPhys.2.1.003}%
  \BibitemOpen
  \bibfield  {author} {\bibinfo {author} {\bibfnamefont {P.}~\bibnamefont {Weinberg}}\ and\ \bibinfo {author} {\bibfnamefont {M.}~\bibnamefont {Bukov}},\ }\href {\doibase 10.21468/SciPostPhys.2.1.003} {\bibfield  {journal} {\bibinfo  {journal} {SciPost Phys.}\ }\textbf {\bibinfo {volume} {2}},\ \bibinfo {pages} {003} (\bibinfo {year} {2017})}\BibitemShut {NoStop}%
\bibitem [{\citenamefont {Weinberg}\ and\ \citenamefont {Bukov}(2019)}]{quspin_10.21468/SciPostPhys.7.2.020}%
  \BibitemOpen
  \bibfield  {author} {\bibinfo {author} {\bibfnamefont {P.}~\bibnamefont {Weinberg}}\ and\ \bibinfo {author} {\bibfnamefont {M.}~\bibnamefont {Bukov}},\ }\href {\doibase 10.21468/SciPostPhys.7.2.020} {\bibfield  {journal} {\bibinfo  {journal} {SciPost Phys.}\ }\textbf {\bibinfo {volume} {7}},\ \bibinfo {pages} {020} (\bibinfo {year} {2019})}\BibitemShut {NoStop}%
\bibitem [{\citenamefont {Hubbard}\ and\ \citenamefont {Flowers}(1964)}]{hubbard-3}%
  \BibitemOpen
  \bibfield  {author} {\bibinfo {author} {\bibfnamefont {J.}~\bibnamefont {Hubbard}}\ and\ \bibinfo {author} {\bibfnamefont {B.~H.}\ \bibnamefont {Flowers}},\ }\href {\doibase 10.1098/rspa.1964.0190} {\bibfield  {journal} {\bibinfo  {journal} {Proceedings of the Royal Society of London. Series A. Mathematical and Physical Sciences}\ }\textbf {\bibinfo {volume} {281}},\ \bibinfo {pages} {401} (\bibinfo {year} {1964})}\BibitemShut {NoStop}%
\bibitem [{\citenamefont {Fulde}(2012)}]{fulde2012electron}%
  \BibitemOpen
  \bibfield  {author} {\bibinfo {author} {\bibfnamefont {P.}~\bibnamefont {Fulde}},\ }\href@noop {} {\emph {\bibinfo {title} {Electron correlations in molecules and solids}}},\ Vol.\ \bibinfo {volume} {100}\ (\bibinfo  {publisher} {Springer Science \& Business Media},\ \bibinfo {year} {2012})\BibitemShut {NoStop}%
\bibitem [{\citenamefont {Fishman}\ \emph {et~al.}(2022)\citenamefont {Fishman}, \citenamefont {White},\ and\ \citenamefont {Stoudenmire}}]{ITensor}%
  \BibitemOpen
  \bibfield  {author} {\bibinfo {author} {\bibfnamefont {M.}~\bibnamefont {Fishman}}, \bibinfo {author} {\bibfnamefont {S.~R.}\ \bibnamefont {White}}, \ and\ \bibinfo {author} {\bibfnamefont {E.~M.}\ \bibnamefont {Stoudenmire}},\ }\href {\doibase 10.21468/SciPostPhysCodeb.4} {\bibfield  {journal} {\bibinfo  {journal} {SciPost Phys. Codebases}\ ,\ \bibinfo {pages} {4}} (\bibinfo {year} {2022})}\BibitemShut {NoStop}%
\bibitem [{\citenamefont {Gutzwiller}(1964)}]{Gutzwiller}%
  \BibitemOpen
  \bibfield  {author} {\bibinfo {author} {\bibfnamefont {M.~C.}\ \bibnamefont {Gutzwiller}},\ }\href {\doibase 10.1103/PhysRev.134.A923} {\bibfield  {journal} {\bibinfo  {journal} {Phys. Rev.}\ }\textbf {\bibinfo {volume} {134}},\ \bibinfo {pages} {A923} (\bibinfo {year} {1964})}\BibitemShut {NoStop}%
\bibitem [{\citenamefont {Brinkman}\ and\ \citenamefont {Rice}(1970)}]{Brinkman_PhysRevB.2.1324}%
  \BibitemOpen
  \bibfield  {author} {\bibinfo {author} {\bibfnamefont {W.~F.}\ \bibnamefont {Brinkman}}\ and\ \bibinfo {author} {\bibfnamefont {T.~M.}\ \bibnamefont {Rice}},\ }\href {\doibase 10.1103/PhysRevB.2.1324} {\bibfield  {journal} {\bibinfo  {journal} {Phys. Rev. B}\ }\textbf {\bibinfo {volume} {2}},\ \bibinfo {pages} {1324} (\bibinfo {year} {1970})}\BibitemShut {NoStop}%
\bibitem [{\citenamefont {Kotliar}\ and\ \citenamefont {Ruckenstein}(1986)}]{KR-PhysRevLett.57.1362}%
  \BibitemOpen
  \bibfield  {author} {\bibinfo {author} {\bibfnamefont {G.}~\bibnamefont {Kotliar}}\ and\ \bibinfo {author} {\bibfnamefont {A.~E.}\ \bibnamefont {Ruckenstein}},\ }\href {\doibase 10.1103/PhysRevLett.57.1362} {\bibfield  {journal} {\bibinfo  {journal} {Phys. Rev. Lett.}\ }\textbf {\bibinfo {volume} {57}},\ \bibinfo {pages} {1362} (\bibinfo {year} {1986})}\BibitemShut {NoStop}%
\bibitem [{\citenamefont {Zhu}\ \emph {et~al.}(2021)\citenamefont {Zhu}, \citenamefont {Li}, \citenamefont {Han},\ and\ \citenamefont {Wang}}]{han_PhysRevB.103.024514}%
  \BibitemOpen
  \bibfield  {author} {\bibinfo {author} {\bibfnamefont {H.-S.}\ \bibnamefont {Zhu}}, \bibinfo {author} {\bibfnamefont {Z.}~\bibnamefont {Li}}, \bibinfo {author} {\bibfnamefont {Q.}~\bibnamefont {Han}}, \ and\ \bibinfo {author} {\bibfnamefont {Z.~D.}\ \bibnamefont {Wang}},\ }\href {\doibase 10.1103/PhysRevB.103.024514} {\bibfield  {journal} {\bibinfo  {journal} {Phys. Rev. B}\ }\textbf {\bibinfo {volume} {103}},\ \bibinfo {pages} {024514} (\bibinfo {year} {2021})}\BibitemShut {NoStop}%
\end{thebibliography}%

\clearpage
\onecolumngrid
\begin{center}
\textbf{\large Supplemental Material: }
\end{center}

\setcounter{equation}{0}
\setcounter{figure}{0}
\setcounter{table}{0}
\setcounter{page}{1}
\makeatletter
\renewcommand{\theequation}{S\arabic{equation}}
\renewcommand{\thefigure}{S\arabic{figure}}
\renewcommand{\bibnumfmt}[1]{[S#1]}
\renewcommand{\citenumfont}[1]{S#1}

\onecolumngrid
\tableofcontents

\section{Many-body Hamiltonian under HK interaction} \label{many-body-hk}
In this section, we give the analytical form of the many-body Hamiltonian under HK interaction of the two models discussed in our work. Using these momentum space localized many-body Hamiltonian, we can calculate the phase diagram, topological index, and Green function explicitly.  
\subsection{Topological nanowire}

 The many-body Hamiltonian of topological nanowire under HK interaction(Eq.\ref{nano-HK-k-ham}) is expanded by Fock basis $\ket{n_{k,\uparrow},n_{-k,\uparrow},n_{k,\downarrow},n_{-k,\downarrow}}$($n_{k,\sigma},n_{-k,\sigma}=0,1$) at $k$-point, and pairing term divide it into even and odd number block. The Hamiltonian in even number block reads

\bea
\label{h_even_nano}
H_{even}^{nano}(k)=\begin{bmatrix}
 -(h_1+h_2) & 0 & 0 & \Delta & \Delta & 0 & 0 & 0\\
 0 & h_1-h_2 & 0 & -h_{soc} & -h_{soc} & 0 & 0 & 0\\
 0 & 0 & U & 0 & 0 & 0 & 0 & 0\\
 \Delta & h_{soc} & 0 & 0 & 0 & 0 & h_{soc} &  \Delta\\
 \Delta & h_{soc} & 0 & 0 & 0 & 0 & h_{soc} &  \Delta\\
 0 & 0 & 0 & 0 & 0 & U & 0 & 0\\
 0 & 0 & 0 & -h_{soc} & -h_{soc} & 0 & h_2-h_1 & 0 \\
 0 & 0 & 0 &  \Delta &  \Delta & 0 & 0 & h_1+h_2+2U
\end{bmatrix}
\eea
which is under the basis $\{\ket{0000},\ket{1100}\ket{1010},\ket{1001},\ket{0110},\ket{0101},\ket{0011},\ket{1111}\}$. $H_{even}(k)$ holds the information of GS. And the Hamiltonian in odd number block reads
\bea
\label{h_odd_nano}
H_{odd}^{nano}(k)=\begin{bmatrix}
 -h_2 & 0 & h_{soc} & 0 & 0 & 0 & 0 &  \Delta\\
 0 & -h_2 & 0 & -h_{soc} & 0 & 0 & - \Delta & 0\\
 -h_{soc} & 0 & -h_1 & 0 & 0 & - \Delta & 0 & 0\\
 0 & h_{soc} & 0 & -h_1 &  \Delta & 0 & 0 &  0\\
 0 & 0 & 0 &  \Delta & h_2+U & 0 & h_{soc} &  0\\
 0 & 0 & - \Delta & 0 & 0 & h_2+U & 0 & -h_{soc}\\
 0 & - \Delta & 0 & 0 & -h_{soc} & 0 & h_1+U & 0 \\
\Delta & 0 & 0 &  0 &  0 & h_{soc} & 0 & h_1+U
\end{bmatrix}
\eea
which is under the basis $\{\ket{1000},\ket{0100}\ket{0010},\ket{0001},\ket{0111},\ket{1011},\ket{1101},\ket{1110}\}$. $H_{odd}(k)$ holds the information of ES. And $h_1=\varepsilon_k+h$, $h_2=\varepsilon_k-h$ and $h_{soc}=i\alpha_{soc}\cdot sin(k)$ in Eq,\ref{h_even_nano} and Eq.\ref{h_odd_nano}.

\subsection{Spinful Kitaev chain} 

The many-body Hamiltonian of spinful Kitaev chain under HK interaction in reciprocal space can be expanded by the same Fock basis as topological nanowire. The Hamiltonian in even block reads

\bea
\label{h_even_kitaev}
H_{even}^{kitaev}(k)=\begin{bmatrix}
 -2\varepsilon(k) & -i\Delta_{\uparrow,k}  & 0 & 0 & 0 & 0 & -i\Delta_{\downarrow,k} & 0\\
 i\Delta_{\uparrow,k} & 0 & 0 & 0 & 0 & 0 & 0 & -i\Delta_{\downarrow,k}\\
 0 & 0 & U & 0 & 0 & 0 & 0 & 0\\
 0 & 0 & 0 & U & 0 & 0 & 0 &  0\\
 0 & 0 & 0 & 0 & U & 0 & 0 &  0\\
 0 & 0 & 0 & 0 & 0 & U & 0 & 0\\
 i\Delta_{\downarrow,k} & 0 & 0 &0 & 0 & 0 & 0 &-i\Delta_{\uparrow,k} \\
 0 & i\Delta_{\downarrow,k} & 0 &  0 &  0 & 0 & i\Delta_{\uparrow,k} & 2\varepsilon(k)+2U
\end{bmatrix}
\eea
And the many-body Hamiltonian under odd number block reads
\bea
\label{h_odd_kitaev}
H_{odd}^{kitaev}(k)=\begin{bmatrix}
 -\varepsilon(k) & 0 & 0 & 0 & 0 & -i\Delta_{\downarrow,k}  & 0 & 0\\
 0 & -\varepsilon(k) & 0 & 0 & -i\Delta_{\downarrow,k}  & 0 & 0 & 0\\
 0 & 0 & -\varepsilon(k) & 0 & 0 & 0 & 0 & -i\Delta_{\uparrow,k} \\
 0 & 0 & 0 & -\varepsilon(k) & 0 & 0 & -i\Delta_{\uparrow,k}  &  0\\
 0 & i\Delta_{\downarrow,k}  & 0 & 0 & \varepsilon(k)+U & 0 & 0 &  0\\
 i\Delta_{\downarrow,k}  & 0 & 0 & 0 & 0 & \varepsilon(k)+U & 0 & 0\\
 0 & 0 & 0 &i\Delta_{\uparrow,k} & 0 & 0 & \varepsilon(k)+U & 0 \\
 0 & 0 & i\Delta_{\uparrow,k} &  0 &  0 & 0 & 0 & \varepsilon(k)+U
\end{bmatrix}
\eea
Where $\varepsilon(k)=-2t\cdot cos(k)-\mu$, $\Delta_{\sigma,k}=(-)^\sigma\cdot\Delta sin(k)$ in Eq.\ref{h_even_kitaev} and Eq.\ref{h_odd_kitaev}.

\section{Real space Hamiltonian}
\label{many-body-real}
In this section, we give the Hamiltonian form in real space of the two models, which are served for numerical calculation in real space, including the solutions to the phase diagram and entanglement spectrum of the Hubbard model, energy gaps, MZMs and etc.

Firstly, two types of interaction in real space reads

\bea
H_{Hub}=U\sum_{j}n_{{j},\uparrow}n_{{j},\downarrow}\;;\;H_{HK}=\frac{U}{L}\sum_{j_1+j_2=j_3+j_4}c^\dagger_{j_1,\uparrow}c^\dagger_{j_2,\downarrow}c_{j_3,\downarrow}c_{j_4,\uparrow}
\eea
While Hubbard interaction is local in real space, HK interaction is long-range.

The non-interacting part of topological nanowire reads
\bea
\label{nano-nonint-ham}
H_{nano}=&\sum_{i,\sigma}-t(c_{i,\sigma}^\dagger c_{i+1,\sigma}+h.c.)+\sum_{i,\sigma}(\mu+2t) c_{i,\sigma}^\dagger c_{i,\sigma}+\sum_{i}h(c_{i,\uparrow}^\dagger c_{i,\uparrow}-c_{i,\downarrow}^\dagger c_{i,\downarrow})+\\
&\sum_{i}\lambda(c_{i,\uparrow}^\dagger c_{i+1,\downarrow}-c_{i,\downarrow}^\dagger c_{i+1,\uparrow}+h.c.)+\sum_{i}\Delta(c_{i,\uparrow}^\dagger c_{i,\downarrow}^\dagger+h.c.)
\eea
Which corresponds to the first term in Eq.\ref{nano-HK-k-ham}. The non-interacting part of the spinful Kitaev chain reads

\bea
H_{Kitaev}=\sum_{i,\sigma}-t(c_{i,\sigma}^\dagger c_{i+1,\sigma}+h.c.)+\sum_{i,\sigma}\mu c_{i,\sigma}^\dagger c_{i,\sigma}+\sum_{i,\sigma}\Delta_\sigma( c_{i,\sigma}^\dagger c_{i+1,\sigma}^\dagger+h.c.)
\eea
Which corresponds to the first term in Eq.\ref{kitaev_hk-ham}.

\section{Exact solutions to many-body topological index in HK model}
\label{entangle-hk}

In this section, we introduce our method to calculate the many-body topological index under HK interaction and present the results.


In our TSC models, the local Hilbert space in reciprocal space is expanded by 16 Fock states $\ket{n_{k,\uparrow},n_{-k,\uparrow},n_{k,\downarrow},n_{-k,\downarrow}}$($n_{k,\sigma},n_{-k,\sigma}=0,1$), and the ground state wavefunction can be written as $\ket{GS}=\prod_k\ket{GS}_k$. Thus, the method to detect non-interacting topological properties can be generalized to our model.

The first method is to obtain the Zak phase by calculating the Wilson Loop of $\ket{GS}_k$ along the 1D BZ, the many-body Zak phase can be expressed as

\bea
Z=log[\prod_k \;_k\braket{GS|GS}_{(k+\delta k)mod(2\pi)} ]=\sum_k log[_k\braket{GS|GS}_{(k+\delta k)mod(2\pi)}]
\eea
Here, $k=\frac{2\pi}{L}j$ ($j\in [1,L]$) and $\delta k=\frac{2\pi}{L}$. The values Zak phase($Z$) of the two models under HK interaction are listed in Table.\ref{table1}.

The second method is to calculate the entanglement spectrum by performing $\ket{GS}_k$ directly to real space. This method has the same spirit as the Zak phase method by mapping the $\ket{GS}_k$ to the Bloch state of a spinless band insulator with 16 internal degrees of freedom, which holds the same topological properties. Thus, we can use the single-particle Green function to extract entanglement entropy. We define the Green function matrix $G$ in reciprocal space as $G_{k;\alpha,\beta}=_{k,\alpha}\braket{GS|GS}_{k,\beta}$, where ($\alpha,\beta$) denotes the 16 internal degree of freedom in $\ket{GS}_k$. By performing Fourier transformation on $G_{k;\alpha,\beta}$, we obtain the Green function matrix in real space
\bea
G_{i\alpha,j\beta} = \sum_{k} \;_{k,\alpha}\braket{GS|GS}_{k,\beta} e^{-ik(i-j)}
\eea
Here, $0<i,j<N$. $N$ is the size of the mapped model in real space and we can set it very large due to the analyticity. 

In order to solve entanglement entropy and spectrum, we need to choose a subspace $A$, which satisfies $N_{A}<A$ and $N_{A}$ is even. Then, the entanglement spectrum $f_{i,\alpha}$ can be obtained by diagonalizing the Green function matrix of subsystem $A$, which is denoted as $G_{sub}=G_{16\times N_A,16\times N_A}$. And the entanglement entropy can be expressed by entanglement spectrum\cite{wu_PhysRevB.91.115118,fan_PhysRevB.94.165167,Sirker_2014,Peschel_2009} as follow 

\bea
S_{von}(A) = -\sum_{i,\alpha} f_{i,\alpha}ln(f_{i,\alpha}) + (1-f_{i,\alpha})ln(1-f_{i,\alpha})
\eea
Where $S_{von}(A)$ is the Von Neumann entropy of subsystem $A$. Note that at the topological transition point, there is a jump change in the entanglement spectrum.

\begin{figure}[H]
    \centering
    \includegraphics[width=1.0\linewidth]{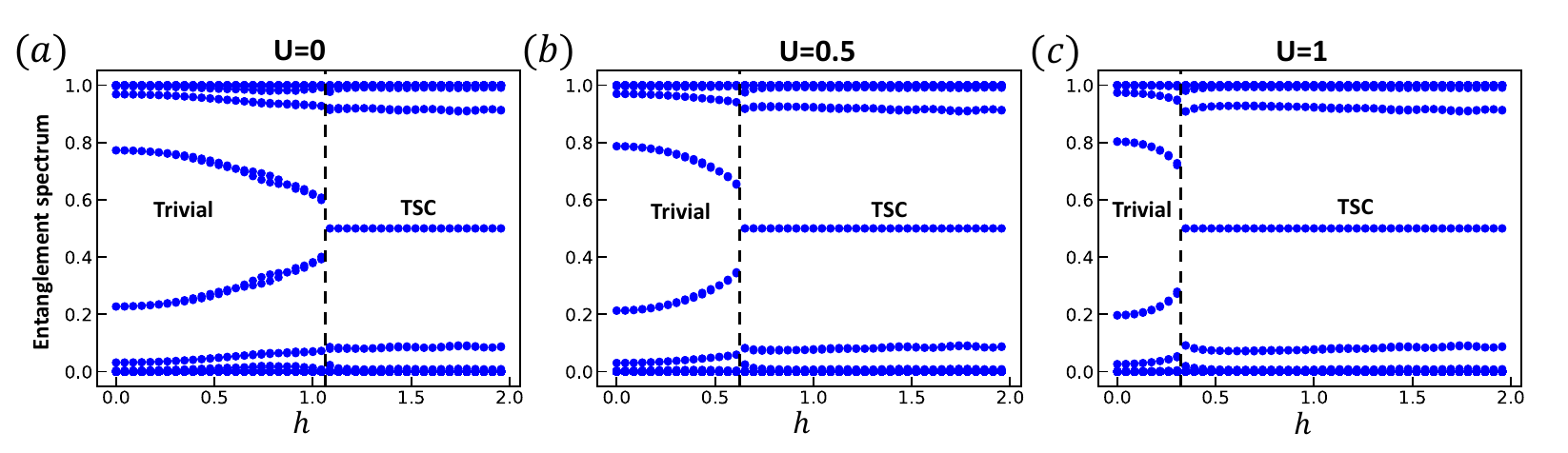}
    \caption{ Entanglement spectrum of the GS in topological nanowire under HK interaction with $U=0$(a), $U=0.5$(b) and $U=1$(c). And the other parameters are $\mu=-1, \lambda=0.5, \Delta=0.4$. TSC and Trivial phase corresponds to those in Flg.\ref{fig2}(a)}
    \label{nano_es}
\end{figure}

\begin{figure}[H]
    \centering
    \includegraphics[width=1.0\linewidth]{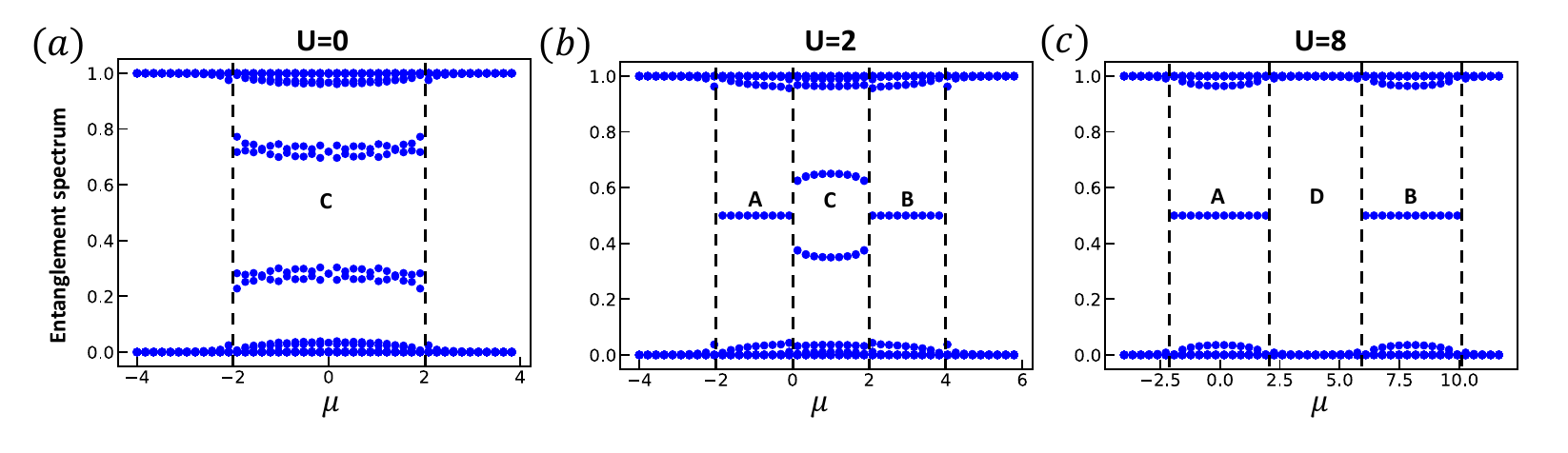}
    \caption{ Entanglement spectrum of the GS in spinful Kitaev chain under HK interaction with $U=0$(a), $U=2$(b) and $U=8$(c). And the other parameters are $t=1, \Delta=0.4$. Region $A$,$B$,$C$ and $D$ corresponds to those in Flg.\ref{fig4}(a). }
    \label{kitaev_es}
\end{figure}

The entanglement spectrum of topological superconductor nanowires under HK interaction with $U=0$, $0.5$ and $U=1$ are shown in Fig.\ref{nano_es}. There are two non-trivial entanglement spectrums in TSC(holon) phase(Fig.\ref{fig2}(a)). And entanglement spectrum of the spinful Kitaev chain under HK interaction with $U=0$, $2$, and $U=8$ are shown in Fig.\ref{kitaev_es}, there are two non-trivial entanglement spectrums in the region $A$ and $B$(Fig.\ref{fig4}(a)) and four non-trivial entanglement spectrum in the region $C$(Fig.\ref{fig4}(a)). The number of non-trivial entanglement spectrums of the two models under HK interaction is listed in Table.\ref{table1}.

\begin{table}[H]
\centering
\begin{tabular}{|c|c|c|c|c|c|c|}
\hline
& \multicolumn{2}{c|}{ nanowire } &\multicolumn{4}{c|}{ Kitaev } \\
\hline
$\;\;\;\;\;\;\;\;\;\;$&$\;\;$Trivial$\;\;$&$\;\;$TSC$\;\;$&$\;\;$A$\;\;$&$\;\;$B$\;\;$&$\;\;$C$\;\;$&$\;\;$D$\;\;$  \\
\hline
Z& $0$ & $\pi$  & $\pi$ & $\pi$ & $0$ & $0$    \\
\hline
N& $0$ & $2$  & $2$ & $2$ & $4$ & $0$ \\
\hline
\end{tabular}
\caption{Zak phase($Z$) and number of non-trivial entanglement spectrum($N$) of different regions in the phase diagrams of topological nanowore(Fig.\ref{fig2}.(a)) and spinful Kitaev chain(Fig.\ref{fig4}.(a)) under HK interaction}
\label{table1}
\end{table}

\section{Numerical solutions to entanglement spectrum in Hubbard model}
\label{entangle-hub}

In this section, we solve the entanglement spectrum in Hubbard model by performing ED. To extract the entanglement spectrum of a finite size chain under PBC, we need to divide the system into two parts(A,B) from the middle, And the reduced density matrix of the subsystem A reads $\rho_A=Tr_B[\rho]$ and can be obtained using ED, where $\rho$ is the density matrix of the whole system. The Von Neumann entropy between A and B has the form of $S_{von} = -Tr[\rho_A ln \rho_A] = -\sum_{i}f_ilnf_i$.
Where $f_i$ is the eigenvalue of $\rho_A$, which is also the entanglement spectrum we want.

\begin{figure}[H]
    \centering
    \includegraphics[width=1.0\linewidth]{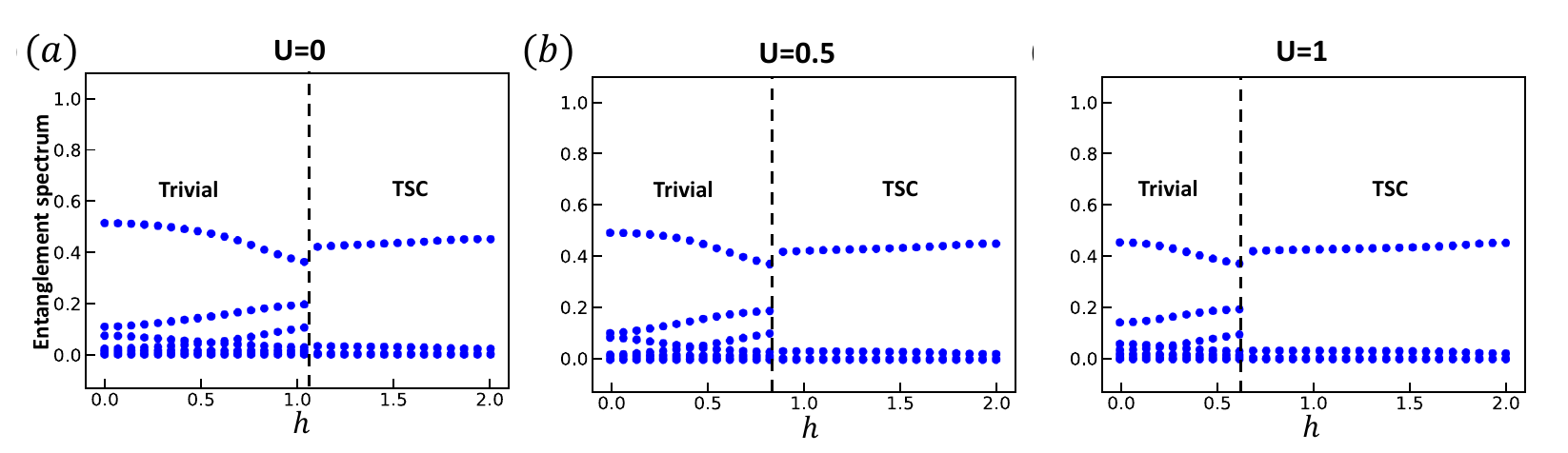}
    \caption{ Entanglement spectrum of the GS in topological nanowire under Hubbard interaction with $U=0$(a), $U=0.5$(b) and $U=1$(c). And the other parameters are $\mu=-1, \lambda=0.5, \Delta=0.4$. TSC and Trivial phase corresponds to those in Flg.\ref{fig2}(b). These results are calculated by performing ED under PBC with 12 sites. }
    \label{nano_es_hub}
    
\end{figure}

\begin{figure}[H]
    \centering
    \includegraphics[width=1.0\linewidth]{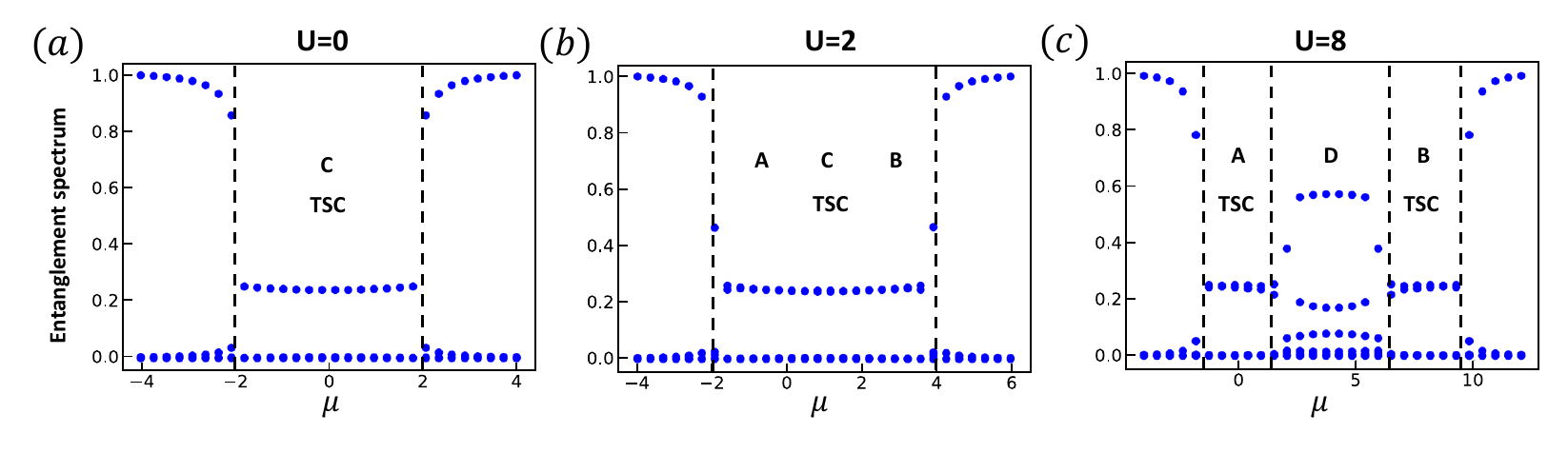}
    \caption{ Entanglement spectrum of the GS in spinful Kitaev chain under Hubbard interaction with $U=0$(a), $U=2$(b) and $U=8$(c). And the other parameters are $t=1, \Delta=0.4$. Region $A$,$B$,$C$ and $D$ corresponds to those in Flg.\ref{fig4}(b). These results are calculated by performing ED under PBC with 12 sites.}
    \label{kitaev_es_hub}
\end{figure}

Entanglement spectrum of topological superconductor nanowire under Hubbard interaction with $U=0$, $0.5$ and $U=1$ are shown in Fig.\ref{nano_es_hub}, there are two non-trivial enetanglement spectrum in TSC(holon) phase(Fig.\ref{fig2}(b)). 
Entanglement spectrum of spinful Kitaev chain under HK interaction with $U=0$, $2$ and $U=8$ are shown in Fig.\ref{kitaev_es_hub}, there are four non-trivial enetanglement spectrum in region A, B and C(Fig.\ref{fig4}(b)), which is different from HK model.

\section{Methodology to solve MZMs under correlation}
\label{mzm_method}

In this section, we introduce our method to extract MZMs under correlation. The idea of solving probability in Eq.\ref{probability} comes from the Green function and low energy local density of state(LDOS)\cite{han_PhysRevB.103.024514}. Similar with electron operator, we define the single particle retarded Green function of majorana operator as $G_{\gamma_{i,\alpha,\sigma};\gamma_{j,\beta,\sigma}}^{R}(t-t') = -i\theta(t-t')<[\gamma_{i,\alpha,\sigma},\gamma_{j,\beta,\sigma}]_+>_0$. After performing Fourier transformation on time, we obtain the Lehmann representation of majorana Green function

\bea
G_{\gamma_{i,\alpha,\sigma};\gamma_{j,\beta,\sigma}}^{R}(\omega)=\sum_n \frac{\bra{0}\gamma_{i,\alpha,\sigma}\ket{n}\bra{n}\gamma_{j,\beta,\sigma}\ket{0}}{\omega-(E_n-E_0)+i0^+} + \frac{\bra{0}\gamma_{j,\beta,\sigma}\ket{n}\bra{n}\gamma_{i,\alpha,\sigma}\ket{0}}{\omega+(E_n-E_0)+i0^+}
\eea
Where $i,j$ denote the lattice index, $\alpha,\beta\in\{1,2\}$ denote the type of majorana fermion and $\ket{0}=\ket{GS}$. To extract the MZMs in real space, we set $\ket{n}=\ket{ES}$, which is the gapless excitation. And we can obtain the local excitation probabilities by solving the spectrum function as follow

\bea
A_{\gamma_{j,\alpha,\sigma}}(\omega)=m_{j,\alpha,\sigma}[\delta(\omega-E_E+E_G)+\delta(\omega+E_E-E_G)]
\eea
Here, $m_{j,\alpha,\sigma}=|\bra{ES}\gamma_{\alpha,j,\sigma}\ket{GS}|^2$ is the probabilities in Eq.\ref{probability}. And the probabilities of projected majorana excitation $\tilde{m}_{j,\alpha,\sigma}$ can be obtained by defining the single particle Green function of projected majorana operator.

Moreover, there are also double MZMs excitation in correlated spinful Kitaev chain(region C in Fig.\ref{fig4}.(a) and region A,B,C in Fig.\ref{fig4}.(b)). To extract double MZMs in real space, we can simply generalize our method to define the double praticle Green function of majorana fermion $G_{\gamma_{j,1,\sigma}\gamma_{j,2,\overline{\sigma}}}^{R}(t-t') = -i\theta(t-t')<[\gamma_{j,2,\overline{\sigma}}\gamma_{j,1,\sigma},\gamma_{j,1,\sigma}\gamma_{j,2,\overline{\sigma}}]_+>_0$, and the local excitation probabilities obtained from spectrum function reads

\bea
\label{mp-probab}
mp_{1,j}=|\bra{ES}\gamma_{j,1,\uparrow}\gamma_{j,2,\downarrow}\ket{GS}|^2 \;;\; mp_{2,j}=|\bra{ES}\gamma_{j,1,\downarrow}\gamma_{j,2,\uparrow}\ket{GS}|^2
\eea

\section{MZMs in correlated spinful Kitaev chain}
\label{mzm_kitaev}

In this section, we give the results of single and double MZMs excitation in the correlated spinful Kitaev chain. 

\subsection{Single MZMs}

\begin{figure}[H]
    \centering
    \includegraphics[width=1.0\linewidth]{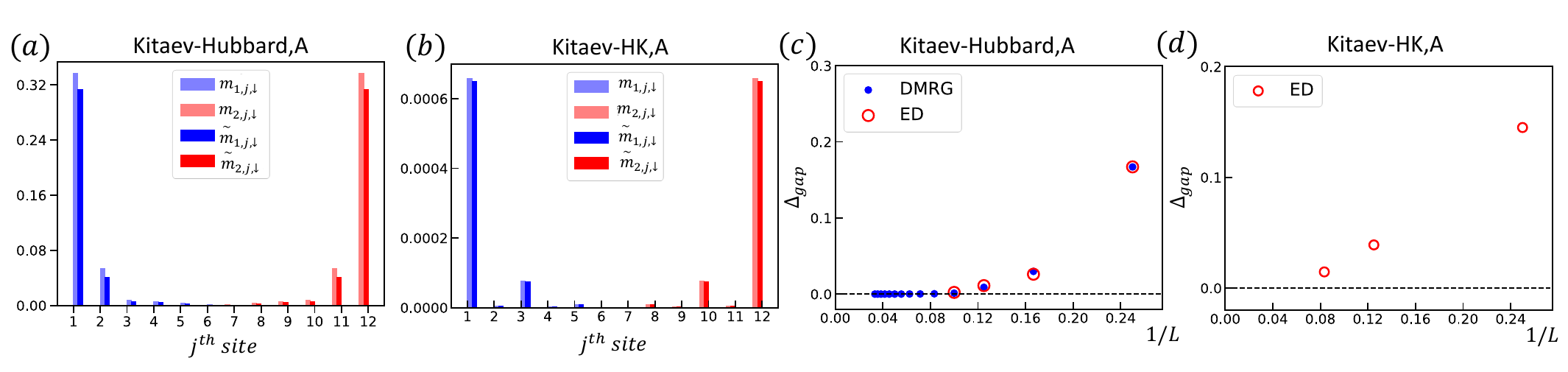}
    \caption{(a),(b) are the distributions of $m_{\alpha,j,\downarrow}$ and $\tilde{m}_{\alpha,j,\downarrow}$ between GS and 1$^{st}$ ES under HK and Hubbard interaction with $\Delta=0.4,U=16,\mu=0$, which is calulated by ED with 12 sites under OBC. (c),(d) shows the gap-closing feature between GS and 1$^{st}$ ES of the model under HK and Hubbard interaction with the same parameters as (a),(b) with different length under OBC. }
    \label{majorana_h}
    
\end{figure}

\begin{figure}[H]
    \centering
    \includegraphics[width=1.0\linewidth]{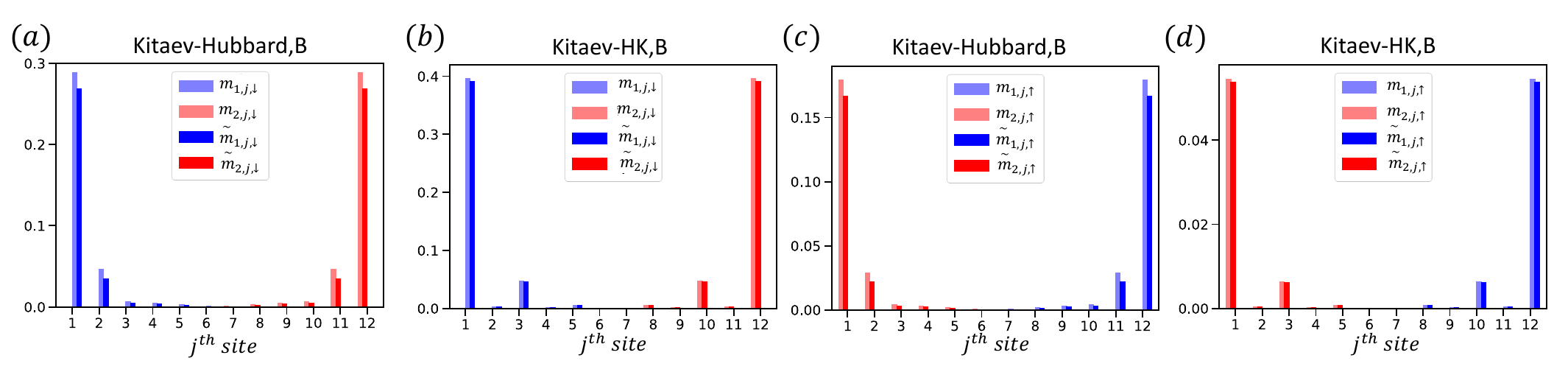}
    \caption{(a),(b) are the distributions of $m_{\alpha,j,\downarrow}$ and $\tilde{m}_{\alpha,j,\downarrow}$ between GS and 1$^{st}$ ES under HK and Hubbard interaction with $\Delta=0.4,U=16,\mu=16$, which is calulated by ED with 12 sites under OBC. (c),(d) are the distributions of $m_{\alpha,j,\uparrow}$ and $\tilde{m}_{\alpha,j,\uparrow}$ with the same other conditions as (a),(b). }
    \label{majorana_d}
    
\end{figure}

To extract the single projected MZMs excitation, we solve the  $m_{j,\alpha,\sigma}$ and  $\tilde{m}_{j,\alpha,\sigma}$ defined in Eq.\ref{probability} under both HK and Hubbard interaction. Region A and B(Fig.\ref{fig4}(a),(b)) are two particle-hole symmetric region with electron projection $(1-n_{\alpha,\sigma})$ and $n_{\alpha,\sigma}$, where $\alpha=i$ under Hubbard interaction and $\alpha=k$ under HK interaction. The results of spin-$\uparrow$ projected MZMs in region A are shown in the main text(Fig.\ref{fig4}.(c),(d)), and the results of spin-$\downarrow$ projected MZMs are shown in Fig.\ref{majorana_h}.(a),(b). We can find that the low-energy excitation are mainly from projected majorana fermion $\tilde{\gamma}_{\alpha,\sigma}$ localized at two edges. Here, $\tilde{\gamma}_{1,\sigma}$ and $\tilde{\gamma}_{2,\overline{\sigma}}$ are excited at the same edge, which is consistent with $\Delta_\uparrow = -\Delta_\downarrow$ in the spinful Kitaev chain. The results of spin-$\sigma$ projected MZMs in region B are shown in Fig.\ref{majorana_d}, which exhibit the similar properties with that in region A.  

The energy gap $\Delta_{gap}$ between the lowest excitation and ground state in region A under OBC under HK and Hubbard interaction are shown in Fig.\ref{majorana_h}.(c),(d), which become gapless under thermodynamic limit
and show similar gap-closing feature with topological nanowire.

\subsection{Double MZMs}
\begin{figure}[H]
    \centering
    \includegraphics[width=1.0\linewidth]{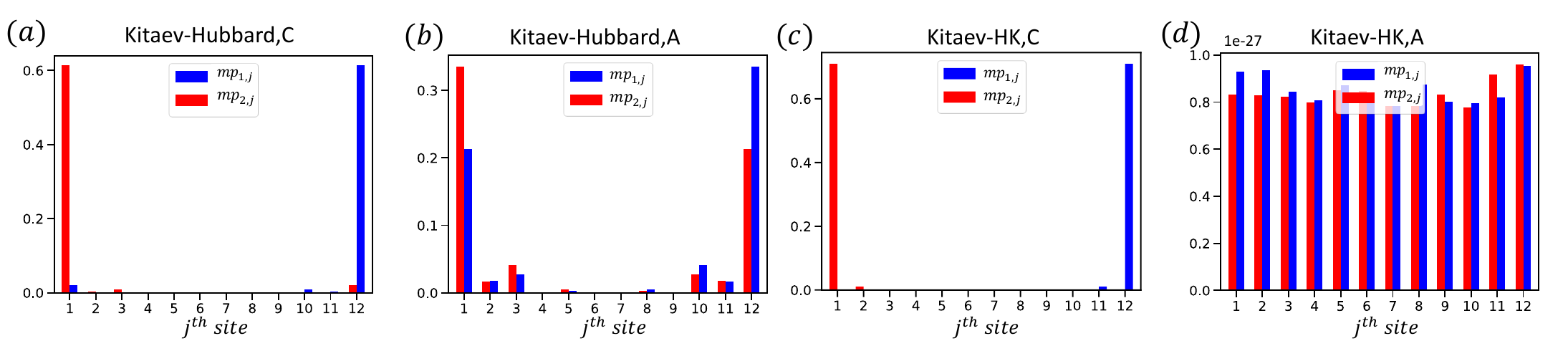}
    \caption{ (a),(b) are the distributions of $mp_{\alpha,j}$ between GS and 3$^{nd}$ ES under Hubbard interaction with $U=1$ and $U=10$, the other parameters are $\Delta=0.4,\mu=0$. (c),(d) are the distributions of $mp_{\alpha,j}$ under HK interaction with the same other conditions as (a),(b). These results are calculated by ED with 12 sites under OBC.}
    \label{majorana_p}
    
\end{figure}

To extract the double projected MZMs excitation, we solve the  $mp_{\alpha,j}$ and defined in Eq.\ref{mp-probab} under both HK and Hubbard interaction in region C and A, as shown in Fig.\ref{majorana_p}. We find that double MZMs is remained under strong Hubbard correlation but eliminated by strong HK correlation. And interestingly, double MZMs excitation will have a hybridization under large Hubbard U, which tends to be the excitations of  $\gamma_{j,1,\uparrow}\gamma_{j,2,\downarrow} + \gamma_{j,1,\downarrow}\gamma_{j,2,\uparrow}$ at two edges, as shown in Fig.\ref{majorana_p}.(b).

\end{document}